\documentclass[12pt,draft]{article}
\usepackage{graphicx}
\usepackage{color}

\makeatletter
\@addtoreset{equation}{section}

\makeatletter
\renewcommand\section{\@startsection {section}{1}{\z@}%
                                   {-3.5ex \@plus -1ex \@minus -.2ex}
                                   {2.3ex \@plus.2ex}%
                                   {\normalfont\large\bfseries}}
\renewcommand\subsection{\@startsection{subsection}{2}{\z@}%
                                     {-3.25ex\@plus -1ex \@minus -.2ex}%
                                     {1.5ex \@plus .2ex}%
                                     {\normalfont\bfseries}}

\def\baselinestretch{1.2}
\newcommand{\be}{\begin{equation}}
\newcommand{\ee}{\end{equation}}
\newcommand{\beq}{\begin{eqnarray}}
\newcommand{\eeq}{\end{eqnarray}}

\makeatletter
\def\@strike{\relax\leavevmode
  \ifmmode
    \expandafter\mathpalette\expandafter\math@strike
  \else
    \expandafter\make@strike
  \fi}
\def\math@strike#1#2{%
  \setbox\z@\hbox{$\m@th#1{#2}$}\fin@strike}
\def\make@strike#1{%
  \setbox\z@\hbox{\color@begingroup#1\color@endgroup}\fin@strike}
\def\fin@strike{%
  \@tempdima\dp\z@
  \@tempdimb\ht\z@
  \lower\@tempdima\hbox{\strike@start}%
  \box\z@
  \raise\@tempdimb\hbox{\strike@end}}
\def\strike@start{\special{ps: %
    currentpoint /starty exch def /startx exch def}}
\def\strike@end{
}
\newcommand\fs{\protect\@strike}

\def\[{\left [}
\def\]{\right ]}
\def\({\left (}
\def\){\right )}

\def\R{{\bf R}}
\def\S{{\bf S}}

\def\e{\varepsilon}

\def\godel{G{\"o}del}

\begin{document}
\begin{titlepage}

\begin{flushright}
hep-th/0405019\\
\end{flushright}

\vfil\

\begin{center}

{\Large{\bf Over-Rotating Black Holes, \godel\ Holography \\ 
\vglue.1in
and the Hypertube}} \vfil

\vspace{3mm}

Eric G. Gimon$^{a}$$^{b}$ and  Petr Ho\v{r}ava$^a$$^{b}$
\\

\vspace{8mm}

$^a$ Department of Physics, University of California,
Berkeley, CA 94720, USA\\

$^b$ Theoretical Physics Group, LBNL, Berkeley, CA 94720, USA

\vfil

\end{center}

\begin{abstract}
\noindent
We demonstrate how a five dimensional G\"{o}del universe appears
as the core of resolved two-charge and three-charge over-rotating BMPV 
black holes. A smeared generalized supertube acts as a domain wall and 
removes regions of closed timelike curves by cutting off both the inside 
and outside solution before causality violations appear, effectively 
allowing the \godel\ universe and the over-rotating black hole to solve 
each other's causality problems.  This mechanism suggests a novel form 
of holography between the compact G\"{o}del region and the diverse vacua 
and excitations of the bound state of a finite number of D0 and D4-branes 
with fundamental strings.
\end{abstract}
\vspace{0.5in}

\end{titlepage}
\renewcommand{\baselinestretch}{1.05}  
\section{Introduction and Summary}

   One of the most productive approaches in theoretical physics is
the development of toy models to extract key physical insight into the 
behavior of related problems which are either technically intractable or very
complex.  One current theoretical problem of great import is the
understanding of de~Sitter spacetimes in the context of a
consistent theory of quantum gravity such as string theory; such
spacetimes are widely believed to describe the very early
universe.  Of particular interest would be a good picture of how
holography applies to de~Sitter space.

    A possible toy model for understanding de~Sitter holography was
proposed by Boyda etal.\ in \cite{BGHV}, where they pointed
out that a supersymmetric \godel\ universe solution
\cite{GGHPR} shared several intriguing properties with de~Sitter
space.  In particular, they showed the existence of observer-dependent 
finite-size preferred holographic screens (in the nomenclature of
\cite{Boussoscreen}) associated with any localized observer in this 
spacetime -- a qualitative feature reminiscent of the holographic 
properties of cosmological horizons of de~Sitter space.  In contrast to 
the puzzling case of the de~Sitter universe, however, the classical \godel\ 
solutions of string theory typically exhibit a large number of Killing 
spinors, raising the hope of using supersymmetry to 
control the behavior of these solutions in full string theory.    

   Unfortunately, while the \godel\ solution is endowed with interesting
holographic properties, it also has closed time-like curves
through every point. Although this spacetime has a globally well-defined
timelike Killing vector (thereby formally allowing supersymmetry), there is
no Cauchy surface, and no manifestly consistent inner product for quantum 
fields, making it difficult to even formulate any preliminary conjecture for a
holographic correspondence between the degrees of the freedom in the bulk
spacetime and the hypothetical holographically dual degrees of freedom on 
any of the locally defined compact screens.  

Another observation in \cite{BGHV}, however, perhaps offers a clue
for further progress. If one selects a single observer in the \godel\
universe, the corresponding preferred screen carves out a tubular region 
of the causality-violating classical \godel\ solution which in itself is 
causally safe.  This led the authors of \cite{BGHV} to conjecture that 
holography could play a crucial role in resolving the causality problems 
in some solutions of string theory, effectively suggesting the possibility 
of ''holographic chronology protection.'' The intriguing structure of 
holographic screens in the \godel\ universe has been 
determined in \cite{BGHV} by purely kinematic arguments, using Bousso's 
``phenomenological'' definition of preferred screens; however, 
no insight into the microscopic theory of such hypothetical screens was 
offered in \cite{BGHV}.  Thus, a natural question emerges: Are there 
{\it dynamical\/} objects in string theory that would play the role of 
the holographic screens in the \godel\ universe?  If this question has 
a positive answer, such objects would be codimension-one domain walls 
in spacetime, and their degrees of freedom should be connected to those 
of the bulk \godel\ universe via holographic duality.  

  In \cite{3Dgodel}, Drukker etal.\ demonstrated that a supertube
\cite{Supertube1,Supertube2} had just the right sort of
world-volume to bound a tubular region in a three-dimensional
\godel\ universe with seven compact directions. Unfortunately,
this solution has a few flaws for our purposes: it is not clear
why the supertube should have a radius small enough for the region
within to exclude CTC's, and the asymptotic metric suffers from
logarithmic divergences.

    In order to use \godel\ as a toy model for gaining an understanding
of de Sitter holography, we would like a domain wall sitting at some 
radius smaller than that at which closed timelike curves appear, and such 
that the outside solution is asymptotically flat (or has some other ``good'' 
asymptotic behavior at infinity, for example, asymptotes to AdS space).  
Asymptotic flatness seems useful, 
since it allows one, for example, to introduce black holes inside
\godel\ \cite{Herdeiro,GH,GodelBH2,BDGO,BD} while simultaneously maintaining a
reference for notions of mass and temperature (or even the conventional 
definition of a horizon), otherwise thorny 
problems in the absence of nice asymptotic behavior at infinity 
(see \cite{GodelBH2}).

    In this paper, we will take as motivation the results of 
\cite{3Dgodel}, and use smeared supertubes to bound the 
five-dimensional \godel\ solution of \cite{GGHPR}.  We find an
outside solution which takes the form of a reduction of the extremal 
rotating BMPV black hole \cite{BMPV1,BMPV2,CveticYoum,CveticLarsen} obtained 
by setting one of the three charges of the BMPV solution to zero.%
\footnote{It is tempting to be somewhat cavalier about the terminology, and 
keep referring to the two-charge solutions or the over-rotating three-charge 
solutions as ``black holes,'' despite the fact that they carry no macroscopic 
Bekenstein-Hawking entropy and actually represent naked singularities  
if taken seriously as classical solutions of low-energy supergravity.}
We will also consider as our outside solution the full 
three-charge rotating BMPV black hole, which requires adding a dust of
wrapped D4-branes to the domain wall supertube.  The
supertube/D4-brane system is U-dual to the well studied D1/D5/KK
system; we will use this to examine the microstates of our
solutions.   Finally, we will propose a novel form of holography
relating the \godel\ bulk degrees of freedom to those of the domain
wall.  We will not provide the full explicit construction, but rather
set out the outline of a future program outside the scope of this
paper.

   Our presentation will be organized as follows.  In Section~2 we
will provide the relevant supergravity solutions for 
the five-dimensional \godel\ universe and its possible
"completions" outside the domain wall.  The Israel matching
conditions provide the crucial input for Section~3, where
we detail the microscopic properties of the domain walls for the
two-charge and three-charge black holes.  We will not be considering
the most general domain walls for sourcing these black holes, but
rather look at those directly relevant to the \godel\ solution.
In Section~4 we will give near-horizon (or near-tube) limits,  while
Section~5 will provide some deeper analysis with comments on the
smearing of the supertubes, the relations to the D1/D5/KK CFT and the
near-tube limit.  We will close with an outline of our ideas for
\godel\ holography and some broader general remarks.  

  Just before this article was posted, another paper appeared on a similar
subject~\cite{Drukker}. 

\section{Creating a Divide: The Inside and Outside SUGRA Solutions}

The starting point for our paper
will be a general supertube IIA supergravity solution
 as described by Emparan, Mateos and Townsend in \cite{Supertube2}:
\begin{eqnarray}
\label{ansatz}
ds_{IIA}^2 &=& - U^{-1}V^{-1/2}(dt - A)^2+U^{-1}V^{1/2} dy^2 + V^{1/2}\sum_{i = 1}^8 (dx^i)^2,
\\
B_2 &=& -U^{-1}(dt - A)\wedge dy + dt\wedge dy,\qquad e^{\Phi} = \, U^{-1/2}V^{3/4}, \nonumber\\
C_1 &=& - V^{-1}(dt - A) + dt,\qquad C_3 = - U^{-1}dt\wedge dy
\wedge A \nonumber.
\end{eqnarray}
It turns out that both the \godel\ solution and the corresponding class of
(the two-charge reduction of) BMPV black holes can both be rewritten in this 
form (or, in the three-charge
case, a slight generalization thereof).  Here $y$ is a coordinate on
$S^1$ of radius $R_9$, and the $x^i$ parametrize the flat $\R^4\times
T^4$, with the volume of $T^4$ denoted by $V_4$.  Also, instead of the flat
coordinates $x^i$, $i=1,\ldots 4$ on the $\R^4$, we will frequently use
spherical coordinates $(r,\phi,\psi,\theta)$, with the conventional Euler
angles $\phi\in[0,4\pi)$, $\psi\in[0,2\pi)$ and $\theta\in[0,\pi)$.
In such spherical coordinates, the functions $U$ and $V$ as well as the 
one-form $A$ (in the basis of invariant Euler one-forms on $S^3$) depend 
on $r$ only.  The ansatz will satisfy the equations of motion if $U$, $V$ 
and $A$ are all harmonic on $\mathbf{R}^4$.

The supertube spacetimes are actually special cases of the IIA
reduced M-theory solutions of Gauntlett, Myers and Townsend~\cite{GMT}
with M5-brane charges set to zero.  The absence of M5-branes means that
the solutions generically preserve eight supersymmetries.
The inside solution and the first outside solution will take
the supertube metric form above.  The inside, a
five-dimensional \godel\ universe, is a
very symmetric supertube spacetime, with twenty supersymmetries.
On the other hand, our second outside solution will require a
slightly more general framework, with one non-zero M5-brane charge
in the M-theory lift and thus only four supersymmetries.

Before we go on to describe these solutions, we would like to
spotlight their symmetries.  We limit our attention to solutions
that respect the $U(1)_L\times SU(2)_R$ subgroup of the maximal possible
rotation symmetry group $SO(4)$ of four space dimensions.  This condition
restricts the angular momentum (or vorticity) tensor to be self-dual.
Our rotation symmetry group acts transitively on 3-spheres, and
therefore suggests a natural class of adapted coordinate systems, in which
the orbits of $U(1)_L\times SU(2)_R$ are squashed 3-spheres of constant
radial coordinate $r$. In such coordinates,  the angular momentum (or
vorticity) of the solutions is captured by the off-diagonal components of
the metric that mix $dt$ with one of the Euler forms on the squashed $S^3$.
We will not discuss the addition of $U(1)_R$ charge until we get to Section~4.

\subsection{The Inside: G\"{o}del Universe}

Using the Euler one-forms $\sigma_i$, $i=1,\ldots 3$ on $S^3$, we can write
the string metric for the basic ${\cal N}=1$ minimal five-dimensional
supergravity \godel\ solution in a manifestly $U(1)_L\times SU(2)_R$ invariant
form, as:

\begin{equation}
 ds_5^2 = - (dt^2 + \beta\,{r^2 \over 2}\, \sigma_3)^2 + dr^2 +
 {r^2 \over 4} (\sigma_1^2 + \sigma_2^2 + \sigma_3^2)
\end{equation}
with a gauge field
\begin{equation}
{\cal A} = -{\sqrt{3} \over 2} A = {\sqrt{3} \over 4}\beta\, r^2
\sigma_3.
\end{equation}

This solution has closed time-like curves.  For example, if we write
$\sigma_3$ as $d\phi + \cos\theta\,d\psi$, then when $g_{\phi\phi}
= {1\over 4} r^2 ( 1 - \beta^2 r^2)$ becomes negative, i.e. when
$\beta r >1$, the orbits of $\partial_\phi$ will become
time-like. On the other hand, it was observed in~\cite{BGHV} that the
portion of spacetime bounded by a wall at any radius $r \le
1/\beta$ is free of causality violations, since any closed timelike curve
in the full spacetime would have to leave the bounded region before returning
to its original point.  A phenomenological approach to holography further
indicated that the natural holographic screens -- as defined by an
observer localized at the origin of the spherical coordinates in the \godel\
geometry -- should be surfaces of some constant value $r=r_H$, strictly
smaller than $1/\beta$.%
\footnote{The effective (super)gravity approach of~\cite{BGHV}
suggested the value of $r_H=\sqrt{3}/2\beta$ for the location of the
preferred screen.  One should expect that in full string theory the value of 
$r_H$ at the location of the ``optimal'' holographic screen can in principle 
undergo a finite shift, with the precise value determined from some 
yet-to-be-developed microscopic theory of holography for the \godel\ solution.}
Thus, placing a domain wall at some
value of $r<1/\beta$ will have the beneficial effect of curing the causality
problems of the classical solution, in a way suggested by the holographic
principle.

We can lift this 5D solution to a 10-dimensional IIA one as
follows \cite{GGHPR}:
\begin{eqnarray}
\label{10Dmetric}
ds_{10}^2 &=& ds_5^2 + dy^2 + \sum_{i = 5}^{8} (dx^i)^2 \\
B_2 &=&  A \wedge dy, \qquad C_3 =  A \wedge (x^5\wedge x^6 -
x^7\wedge x^8) \nonumber
\end{eqnarray}
To get to the supertube form of this metric we T-dualize along the
$x^5x^6$ plane, and then we dualize the corresponding 5-form
potential to a 3-form potential ($A$ is self-dual in the 1234
subspace).  This leaves the NS-NS fields unchanged, while the
RR fields now become
\begin{equation}
C_1 = A, \qquad C_3 = - dt \wedge dy \wedge A.
\end{equation}
Clearly, this Type IIA spacetime belongs to the class described by our main
ansatz (\ref{ansatz}), with the choice of functions $U=1$, $V=1$, $A= -{1\over
2} r^2 \sigma_3$.  In this form this solution is a
dimensional reduction of the M-theory plane-wave with 24
supercharges, as emphasized in \cite{HT}.

We will use this lift to M-theory to briefly review the supersymmetries.
The metric in eleven dimensions can be written as
 \be
ds^2_{11} = 2e^+e^- + e^ye^y + e^ie^i
 \ee
with 1-forms
 \be
 e^+ = dx^+,\quad e^- = - (dx^- - A),\quad e^y = dy,\quad e^i =
 dx^i.
 \ee
The M-theory four-form is simply
 \be
  F_4 = - dx^+\wedge dy\wedge dA.
 \ee
In this basis, it is convenient to define three projectors
 \be
 \Gamma_-,\qquad {1\over 2}\,dA|_{ij}\,\Gamma^i\Gamma^j \equiv
 dA\cdot\Gamma,\qquad\textrm{and}\qquad {1\over 2}\,(1 - \Gamma_y).
 \ee
We will label a basis of 32 spinors in blocks of four,
$\e^{\pm\pm\pm}$, with a ``$+$'' if they are projected out by the
respective projector, and with a ``$-$'' otherwise.  The Killing
spinor equations require that spinors either be projected out by
$\Gamma_-$ or $dA\cdot\Gamma$, so this M-theory solution has 24 Killing
spinors. The spinors which satisfy $\Gamma_-\,\e =0$,
$\e^{+\pm\pm}$, give sixteen standard~\cite{clp} supersymmetries
while the remaining eight, $\e^{-+\pm}$, are supernumerary.

   Of the twenty-four spinors, sixteen are constant:
 \be
  \e^{+++},\qquad \e^{++-},\qquad \e^{+-+},\qquad
  \textrm{and}\qquad \e^{-+-}.
 \ee
The spinors $\e^{-++}$ have non-trivial $x^i$-dependence, while
the spinors $\e^{+--}$ depend on $x^+$.  The latter dependence
means that we reduce the M-theory plane-wave to our \godel\
solution, only 20 supercharges are preserved.

\subsection{The Outside Part I: The Two-Charge Reduction of the BMPV 
Black Hole}

The 10D IIA supergravity solution for a rotating two-charge reduction of the 
BMPV black hole~\cite{BMPV1,BMPV2},
with D0 and F1 charges $Q_0$ and $Q_s$ respectively,
also takes the form of a supertube metric with eight
supersymmetries:
\begin{eqnarray}
\label{BMPV}
 ds^2 &=& - \tilde{U}^{-1}\tilde{V}^{-1/2}(d\tilde{t} + {J\over 4\tilde{r}^2}\,
 \sigma_3)^2+ \tilde{U}^{-1}
 \tilde{V}^{1/2} d\tilde{y}^2 + \tilde{V}^{1/2}\sum_{i = 1}^8 (d\tilde{x}^i)^2, \\
 B_2 &=& -\tilde{U}^{-1}(d\tilde{t} + {J\over 4\tilde{r}^2}\,\sigma_3)\wedge d\tilde{y} +
 d\tilde{t}\wedge d\tilde{y}, \qquad e^{\Phi} =
 \tilde{g_s}\, \tilde{U}^{-1/2}\tilde{V}^{3/4}, \nonumber\\
 \tilde{g_s}C_1 &=& - \tilde{V}^{-1}(d\tilde{t} + {J\over 4\tilde{r}^2}\,\sigma_3) +
 d\tilde{t}, \qquad \tilde{g_s}C_3 =  \tilde{U}^{-1}d\tilde{t}\wedge
 d\tilde{y} \wedge {J\over 4\tilde{r}^2}\,\sigma_3 \nonumber
\end{eqnarray}
with
\begin{equation}
 \tilde{U}(\tilde{r}) = 1 + {Q_s\over \tilde{r}^2},\qquad \tilde{V}(\tilde{r}) = 1 +
 {Q_0\over \tilde{r}^2}.
\end{equation}
These fields are written in the string frame; we find it
convenient to exhibit $\tilde{g_s}$-dependence in the RR-fields.
The eight supersymmetries are those of all supertube spacetimes,
with $\tilde{r}$-dependent Killing spinors proportional to
$\e^{+-+}$ and $\e^{+++}$~\cite{Supertube2}.

This solution has two pathologies. First, the dilaton-gradient
and metric-curvature diverge at $\tilde{r} =0$. Second, like the
\godel\ solution, this metric also develops CTC's: orbits of the
vector field $\partial_\phi$ with radii smaller than a
critical value $\tilde{r}_c$ satisfying
 \begin{equation}
 \label{rc}
  4\,\tilde{r}_c^2 \,(\tilde{r}_c^2 + Q_0^2)\, (\tilde{r}_c^2 +
  Q_s^2) =  J^2
 \end{equation}
provide a clear example.  In fact, if we restrict ourselves to the
region {\em outside} this critical radius, there are no CTC's. We
now see that an appropriate match of the inside solution (\godel)
with the outside solution (BMPV) could remove all closed
time-like curves (as well as the pathological throat region).

    We are interested in matching this outside solution to our
inside \godel\ solution at a fixed radius $\tilde{R}$. At that
point, the harmonic functions take values
\begin{equation}
  \chi_s \equiv \tilde{U}(\tilde{R}) = 1 + {Q_s\over \tilde{R}^2} ,
 \qquad \chi_0 \equiv \tilde{V}(\tilde{R}) = 1 + {Q_0\over \tilde{R}^2}.
\end{equation}
In order for the dilaton and the metric to remain real, we also
require $\chi_s,\chi_0 \ge 1$, i.e. we will only consider putting
our domain wall at positive $\tilde{R}^2$.  This restriction is of course 
implied by the fact that $\tilde{r}^2=0$ is a singularity.  

When we cut off the outside solution at any finite value
$\tilde{R}$ of the $\tilde{r}$ coordinate and try to match it with
the G\"{o}del solution, the
$\tilde{t},\tilde{r},\tilde{x}^i,\tilde{y}$ coordinate system
does not represent a smooth continuation of the coordinate system that 
we used in the previous subsection to describe the \godel\ solution.  
For example, the inside solution
always has $g_{rr}=1$ while for the outside solution the value of
$g_{\tilde{r}\tilde{r}}$ at the fixed cut-off radius is $\sqrt{\chi_0}$.  
This discrepancy is an artifact of using a coordinate system which is not 
smoothly extended across the domain wall.  The correct 
way to define what is a smooth coordinate system is to demand the equality of
{\it all\/} the metric components at the wall, not just of the metric induced
on the wall.  Two natural smooth coordinate systems suggest themselves:

(1) One can extend the $(\tilde{t},\tilde{r},\tilde{x}^i,\tilde{y})$
coordinate system from the outside to the region inside the domain wall.  This
requires specific constant rescalings of the original \godel\ coordinates.
The advantage of this coordinate system is that the metric now asymptotes to
the canonical Minkowski metric at infinity, at the cost of introducing
inconvenient rescaling constants in the \godel\ region at the core
of the solution.

(2) Alternatively, one can keep the original \godel\ coordinates
$(t,r,x^i,y)$ on the inside, and extend them to the outside region by constant
rescalings of the original outside coordinates.  We find that this coordinate
system is better suited for the description of the region near the domain
wall, at the price of losing the nice asymptotic property of the metric
components at infinity.  We will refer to this coordinate system as the
``near-wall'' or ``near-tube'' coordinate system.

The rescaling between the two coordinate systems takes the
following form:
\begin{equation}
\label{neartubeI}
\tilde{t} = \chi_s^{1/2} \chi_0^{1/4}\, t,\qquad \tilde{y}=
\chi_s^{1/2} \chi_0^{-1/4}\, y, \qquad \tilde{x}^i =
\chi_0^{-1/4}\, x^i .
\end{equation}

  In order to match at the domain wall, both geometries must induce the
same metric on the wall itself.  To match the volumes of $T^4$, $S^1$ and 
$S^3$, as well as the squashing of the $S^3$, we require that
\be
 \tilde{V_4} = \chi_0^{-1}V_4,\;\;
 \tilde{R_9} = \chi_s^{1/2} \chi_0^{-1/4}R_9,\;\;
 \tilde{R}   = \chi_0^{-1/4} R,\;\;
 J = \chi_s^{1/2} \chi_0^{-1/4}\, 2\beta\! R^4.
\ee
The string coupling and the six-dimensional Newton constant
at the domain wall
are
\begin{equation}
\tilde{g}_s = g_s \chi_s^{1/2} \chi_0^{-3/4},\qquad \tilde{G}_6 =
G_6  \chi_s \chi_0^{-1/2}
\end{equation}

If we place our domain wall at a radius $\tilde{R}>0$ such that we
cut-off our space time before the dilaton gets singular, the
$\chi$'s take values in the interval $[1,\infty)$. It is
convenient to define new parameters
 \be
 \gamma_s =  1 - \chi_s^{-1},\qquad \gamma_0 = 1 - \chi_0^{-1}
 \ee
which take values in the interval $[0,1)$.  In terms of our new
parameters $\gamma$ and the new coordinates,  we can now write the outside
BMPV solution using the ansatz (\ref{ansatz}) with the following choice
of $U, V$ and $A$:
\begin{equation}
U = 1 + {\gamma_s} ({R^2\over r^2}-1), \qquad V = 1 + {\gamma_0}
({R^2\over r^2}-1), \qquad A = - {\beta \over 2} {R^4\over r^2}\,
\sigma_3\,.
\end{equation}
This matching also requires a simple gauge transformation on the potentials.

Near the domain wall, it is sensible to express the
local charges of the BPS algebra in terms of the radius $R$.  For
example, the central charges corresponding to D0-branes and the
$F$-strings are $\gamma_0 R^2$ and $\gamma_s R^2$ respectively,
which can be clearly seen by considering the integer charges as
functions of moduli times the central charges:
\begin{eqnarray}
 n_0 &=&  Q_0 \cdot  \Big(\tilde{g}_s{\pi^2\over 2}\,(\alpha')^{1/2}
                           \tilde{R}_9 \tilde{G}_6^{-1}\Big)
    \,=\, \gamma_0 {R}^2\cdot  \Big({g}_s{\pi^2\over 2}\,
    (\alpha')^{1/2}{R}_9{G}_6^{-1}\Big)
   \nonumber \\
   n_s &=&  Q_s\,{\pi^2 \alpha' \over 2\tilde{G}_6} \,
      =\, \gamma_s {R}^2\cdot \Big( {\pi^2 \alpha'  \over 2{G}_6}\Big).
\end{eqnarray}
This redshift of the central charges plays an important role in
the attractor phenomena of BPS black holes~\cite{ferrara}, but
since we always cut off our solution at non-zero $R$ the charges
never reach any fixed points.

\subsection{The Outside Part II: The Three-Charge BMPV Rotating Extremal Black Hole}

The two-charge solution represents the minimal asymptotically flat solution 
that can serve as a causal outside extension of the \godel\ universe, with 
-- as we will see in the next subsection -- a simple smeared supertube 
playing the role of the domain wall between the two geometries.  
Things become even more interesting when we consider outside solutions with 
more charges, such as the three-charge BMPV black hole.  The 
matching will still be possible, and the domain wall required in the process 
will exhibit some interesting novel properties.  

In this subsection, we start with the following, slightly more generic 
candidate for the outside metric, 
\begin{eqnarray}
\label{BMPV3}
 ds^2 &=& - \tilde{U}^{-1}\tilde{V}^{-1/2}\tilde{W}^{-1/2}(d\tilde{t} + {J\over 4\tilde{r}^2}\,
 \sigma_3)^2+ \tilde{U}^{-1} \tilde{V}^{1/2}\tilde{W}^{1/2}
 d\tilde{y}^2  \\
 &&\qquad\qquad\qquad +\; \tilde{V}^{1/2}\tilde{W}^{1/2}\sum_{i = 1}^4 (d\tilde{x}^i)^2
 + \tilde{V}^{1/2}\tilde{W}^{-1/2}\sum_{i = 5}^8 (d\tilde{x}^i)^2, \nonumber\\
 B_2 &=& -\tilde{U}^{-1}(d\tilde{t} + {J\over 4\tilde{r}^2}\,\sigma_3)\wedge d\tilde{y} +
 d\tilde{t}\wedge d\tilde{y}, \qquad e^{\Phi} =
 \tilde{g_s}\, \tilde{U}^{-1/2}\tilde{V}^{3/4}\tilde{W}^{-1/4}, \nonumber \\
 \tilde{g_s}C_1 &=& - \tilde{V}^{-1}(d\tilde{t} + {J\over 4\tilde{r}^2}\,\sigma_3) +
 d\tilde{t}, \nonumber \\
 \tilde{g_s}C_3 &=&
 {1\over4}\, Q_4 \cos\theta\, d\phi\wedge d\psi\wedge d\tilde{y} + \tilde{U}^{-1}d\tilde{t}\wedge
 d\tilde{y} \wedge {J\over 4\tilde{r}^2}\,\sigma_3 \nonumber
\end{eqnarray}
with
\begin{equation}
 \tilde{U}(\tilde{r}) = 1 + {Q_s\over \tilde{r}^2},\qquad \tilde{V}(\tilde{r}) = 1 +
 {Q_0\over \tilde{r}^2},\qquad \tilde{W}(\tilde{r}) = 1 + {Q_4\over \tilde{r}^2}.
\end{equation}
As mentioned before, this three-charge BMPV solution is a special case of 
the solutions~\cite{GMT} of Gauntlett, Myers and Townsend with four
supersymmetries; we have now set only one of their M5-brane
charges to zero.  The sign of the additional D4-branes
(descendants of the added M5-branes) determines which of the
$\tilde{r}$-dependent Killing spinors remain, those proportional
to $\e^{+-+}$ or those proportional to $\e^{+++}$, thus reducing the number
of supersymmetries.

The addition of the third charge to the metric has an important effect on 
the physics of the solution. The critical radius for CTC's, with a zero-area 
three-sphere, is now located at
\begin{equation}
  4\,(\tilde{r}_c^2 + Q_4) \,(\tilde{r}_c^2 + Q_0)\, (\tilde{r}_c^2 +
  Q_s) =  J^2,
\end{equation}
with $\tilde{r}^2_c$ positive if $J^2 > 4Q_4Q_0Q_s$,
and negative otherwise.  The case of  $J^2 > 4Q_4Q_0Q_s$ describes an
``over-rotating black hole''~\cite{BMPV1,BMPV2,Gibbons}, which in fact is not
a black hole at all; instead, $\tilde r=0$ is a naked singularity, surrounded
by a naked region of CTCs.  This case is very similar to the spinning
two-charge solution that we discussed in the previous subsection.
If $J^2 \leq 4Q_4Q_0Q_s$, the solution describes an honest
black hole whose horizon is located at $\tilde{r}=0$.  The region with CTCs
still exists, but is now cloaked by the horizon, which  is a
squashed three-sphere with a finite sized horizon area
\begin{equation}
{\cal A}_H = \pi^2 (Q_0Q_4)^{1/4}Q_s^{-{1/2}}
\sqrt{4Q_4Q_0Q_s - J^2}.
\end{equation}

Once again we will need to perform a coordinate transformation to
near-tube coordinates which match with the inside solution. Consider placing
the domain wall at $\tilde{r} = \tilde{R}$, (i.e., outside the horizon if
the solution is under-rotating).  To match the coordinate systems, we define
parameters $\chi_i$, much like in the two-charge case:
\begin{equation}
  \chi_s \equiv \tilde{U}(\tilde{R}) = 1 + {Q_s\over \tilde{R}^2} ,
  \qquad \chi_0 \equiv \tilde{V}(\tilde{R}) = 1 + {Q_0\over \tilde{R}^2},
  \qquad \chi_4 \equiv \tilde{W}(\tilde{R}) = 1 + {Q_4\over \tilde{R}^2}.
\end{equation}
Also, just as in the two-charge case, we define new parameters $\gamma_i$
 \be
\label{gmas}
 \gamma_s =  1 - \chi_s^{-1},
 \qquad \gamma_0 = 1 - \chi_0^{-1}
 \qquad \gamma_4 = 1 - \chi_4^{-1}.
 \ee
All $\gamma_i$ belong to $[0,1]$ in order to ensure that the domain wall is
placed outside or at the horizon.

Unlike in the two-charge case, however, we have to take into account the
possibility of putting our domain wall behind the horizon. Even though
$\tilde r$ itself is no longer a good coordinate behind the
horizon, $\rho\equiv\tilde r^2$ is a good function there.  In fact, $\rho$ 
itself can serve as a coordinate that replaces $\tilde r$ behind the horizon;
another particularly nice coordinate that extends through the horizon is 
the radius $r'$ of the three-sphere in the string frame, 
$r'\equiv\tilde r\tilde V^{1/2}\tilde W^{1/2}$.  In the region behind
the horizon, our function $\tilde{r}^2$ (as a function of either $\rho$ or 
$r'$) takes negative values bound from below by the value of the
smallest of the three charges (which we denote by $\widehat Q$),
 \be \tilde{r}^2 \geq -\widehat Q\equiv -\,\textrm{min}(Q_0,Q_4,Q_s),   
 \ee
with $\tilde{r}^2=-\widehat Q$ describing the singularity.  

The possibility of placing the domain wall behind the horizon
(but above the singularity) now extends the range of our original $\gamma$ 
parameters given in (\ref{gmas}) to satisfy
\be
\gamma_i\leq {Q_i \over Q_i- \widehat Q.}
\ee
If the domain wall is placed behind the horizon, the matching of the induced
metric against the \godel\ interior requires the opposite sign of 
the \godel\ vorticity parameter $\beta$.  This is in accord with the 
intuition that the region behind the horizon of the BMPV black hole 
contributes a negative 
amount to the total (positive) value of the angular momentum of the black 
hole.  Since one of the main points of this paper is the possibility of 
resolving {\it naked\/} closed timelike curves, we will restrict our 
attention for the rest of the paper to the domain walls placed above the 
horizon, i.e., we restrict $\gamma_i\in[0,1]$, and $\tilde {R}^2\geq 0$.

Similarly to the two-charge case, we define a smooth coordinate
system across the domain wall using simple rescalings,
 \begin{eqnarray}
\label{neartube2}
&&
\tilde{t} = \chi_s^{1/2} \chi_0^{1/4}\chi_4^{1/4}\, t, \qquad\qquad
\tilde{y}= \chi_s^{1/2} \chi_0^{-1/4}\chi_4^{-1/4}\, y,  \\
&&
\tilde{x}^i|_{i=1..4} = \chi_0^{-1/4}\chi_4^{-1/4}\, x^i, \qquad
\tilde{x}^i|_{i=5..8} = \chi_0^{-1/4}\chi_4^{+1/4}\, x^i. \nonumber
\end{eqnarray}
Matching the volumes of $T^4$, $S^1$ and $S^3$, as well as the squashing of
the $S^3$, gives
\be
 \tilde{V_4} = \chi_0^{-1}\chi_4 V_4,\;\;
 \tilde{R_9} = \chi_s^{1/2} \chi_0^{-1/4} \chi_0^{-1/4}R_9,\;\;
 \tilde{R}   = \chi_0^{-1/4}\chi_4^{-1/4} R,\;\;
 J = \chi_s^{1/2} \chi_0^{-1/4}\chi_4^{-1/4}\, 2\beta\!  R^4.
\ee
The asymptotic string coupling and six-dimensional Newton's
constant in terms of their values at the domain wall are
\begin{equation}
\tilde{g}_s = g_s \chi_s^{1/2} \chi_0^{-3/4} \chi_4^{1/4},\qquad \tilde{G}_6 =
G_6  \chi_s \chi_0^{-1/2} \chi_4^{-1/2}.
\end{equation}
In the near-tube coordinates, the metric now looks like eq.~(\ref{BMPV3}) but
with new harmonic functions in the metric:
\begin{equation}
\label{3harms}
 U = 1 + {\gamma_s} ({R^2\over r^2}-1), \quad V = 1
+ {\gamma_0} ({R^2\over r^2}-1), \quad W = 1 +
{\gamma_4}({R^2\over r^2}-1), \quad A = - {\beta \over 2}
{R^4\over r^2}\, \sigma_3\,.
\end{equation}
We use a series of simple gauge transformations to match the gauge
potentials at $r= R$ with those of the \godel\ solution.

The charges of the BPS algebra at the domain wall are now
$\gamma_0R^2,\gamma_sR^2$ and $\gamma_4R^2$ as can be seen, for
example, by comparing expressions for the number of D4-branes at
infinity and at the tube
\begin{equation}
\label{n4}
n_4 = Q_4\cdot \tilde{g}_s^{-1}\,
  (\alpha')^{-3/2} \tilde{R}_9
    \,=\, \gamma_4 {R}^2\cdot {g}_s^{-1}\,  (\alpha')^{-3/2}  {R}_9.
\end{equation}

If we want to place our domain wall at the horizon, we must
take $\tilde{R}\to 0$ and thus all the $\gamma_i \to 1$.  As
expected, all the central charges converge to same fixed
value~\cite{ferrara}.  Note that in this case, the transformation between 
our two coordinate systems -- as given in eq.~(\ref{neartube2}) -- becomes 
singular, reflecting the infinite redshift from the domain wall to 
asymptotic infinity.  We can still define the near-tube coordinate 
system, though, by extracting the value for $R^2$ from the 
attractor for the central charges of the BPS algebra.  We then get 
$\beta$ from
\begin{equation}
\beta = {1\over 2}\, J R^{-4}, 
\end{equation}
and write the metric for a domain wall at the horizon using the
harmonic forms in eq.~(\ref{3harms}).  In Section~\ref{limit},
we will see how this limit on the parameters $\gamma_i$ is the
U-dual of the Maldacena limit~\cite{Maldacena} of the D1/D5/KK system.

\section{Bridging the Divide: the Domain Wall}

 Now that we have two outside solutions in hand which we would like to match to
the five-dimensional \godel\ universe, we need to see what kind
of domain wall stress-energy the Israel matching conditions will
impose on us. We will then demonstrate for each case how a
particular smeared supertube, along with possible D4-branes,
satisfies this conditions, and match all the relevant parameters.

\subsection{Matching the Solutions}

Given the inside and outside solutions in the previous section,
let us compute the Israel matching conditions through a domain
wall at $r = R$ using the perpendicular unit norm co-vector:
\begin{equation}
\hat{n}_{\nu} dx^{\nu} = \sqrt{g_{rr}} dr.
\end{equation}
The extrinsic curvature is easy to compute since there are no
$g_{r\mu}$ cross-terms in any of the metrics:
\begin{equation}
K_{\mu\nu} = \nabla_{\mu}\hat{n}_{\nu} = {1\over
2}\,(g_{rr})^{-{1\over2}}\,
\partial_r g_{\mu\nu}.
\end{equation}
For the metric (\ref{10Dmetric}) this gives:
\begin{equation}
K^-_{\mu\nu} dx^{\mu}dx^{\nu}|_{r=R} = - (\beta  R\, \sigma_3)(dt
+
\beta\,{R^2 \over 2}\, \sigma_3) + {R\over 4}(\sigma_1^2 +
\sigma_2^2 + \sigma_3^2)
\end{equation}
The trace of this extrinsic curvature is:
\begin{equation}
K^-|_{r=R} = h^{\mu\nu}\,K^-_{\mu\nu}|_{r=R} = {3\over R}
\end{equation}

For the outside solutions we need to be careful, as the metrics so far 
have been string metrics.  The Israel matching condition involves the Einstein
metric which for the first outside solution (with two charges) is:
\begin{equation}
ds_E^2 = - U^{-3/4}V^{-7/8}(dt - A)^2+U^{-3/4}V^{1/8} dy^2 +
U^{1/4}V^{1/8}\sum_{i = 1}^8 (dx^i)^2.
\end{equation}
This gives an extrinsic curvature at $r=R$ of the form (keeping in
mind $U(R)=V(R)=1$):
\begin{eqnarray}
&& \hspace{-25pt}  K^+_{\mu\nu}\, dx^{\mu}dx^{\nu}|_{r=R} =
(3/8\,U' + 7/16\,V')(dt - A)^2   +  \beta R\, \sigma_3(dt -
A)\nonumber \\&&
  +\,(-3/8\,U' + 1/16\,V')\,dy^2 +\, (1/8\,U' + 1/16\,V') \sum_{i = 1}^8 (dx^i)^2 \nonumber \\&&
+ {R\over 4}(\sigma_1^2 + \sigma_2^2 + \sigma_3^2).
\end{eqnarray}
which has trace ($\sigma_3 (dt - A)$ is traceless)
\begin{eqnarray}
K^+|_{r=R} = h^{\mu\nu}\,K^+_{\mu\nu} &=&
(-3/8\,U' - 7/16\,V') + (-3/8\,U' + 1/16\,V') \\
 &+& 7 (1/8\,U' + 1/16\,V')\nonumber + {3\over R}\\
 &=&  (1/8\,U' + 1/16\,V') + {3\over R} \nonumber.
\end{eqnarray}
The Israel matching condition now tells us that the domain wall
stress tensor has to be:
\begin{eqnarray}
\label{ST}
 (8\pi G_{10})\, S_{\mu\nu}^I &=& \gamma_{\mu\nu} dx^{\mu}dx^{\nu} \nonumber \\
 &=& \Big(K^+_{\mu\nu} - K^-_{\mu\nu} - h_{\mu\nu}(K^+ - K^-) \Big)\, dx^{\mu}dx^{\nu} \\
 &=& {1\over 2}(U' +  V')(dt - A)^2 - {1\over 2}U'\,dy^2 +2 (\beta
 R\, \sigma_3)(dt + \beta\,{R^2 \over 2}\,
 \sigma_3) \nonumber \\
 &=&  -{{\gamma_s} + {\gamma_0} \over R}\, \left(dt + {\beta R^2 \over 2}
\sigma_3\right)^2+2\beta R\left(dt + {\beta R^2 \over 2}\sigma_3\right)
\sigma_3  + {\gamma_s \over R}\, dy^2. \nonumber
\end{eqnarray}
A similar set of computations ensures that the jumps in the field strengths 
associated with the $B$-field and the $C$-fields (let's call them 
$\delta H_3$, $\delta G_2$ and $\delta G_4$) is consistent with the charges of
the domain wall.  Here, we will illustrate just one of them, the other 
matchings being a straightforward extension of the same exercise.  Thus, 
the correct definition for the four-form field strength is:
\begin{equation}
G_4 = dC_3 - dB_2\wedge C_1 = - \,U^{-1}V^{-1}\,dy\wedge dA \wedge
(dt - A).
\end{equation}
The equation of motion for $G_4$ is:
\be
d*G_4 + H\wedge G_4 = 2\kappa^2 J_7
\ee
When integrated across the domain wall, the $H\wedge G_4$ term has
insufficient derivatives (i.e., only step-function distributions, no delta
functions) to contribute and we are left with:
\be
\delta(*G_4) = 16\pi\, G_{10}\, {\!\!\delta \over \delta C_3}\, S_{DW}.
\ee
An explicit evaluation of the jump $\delta (*G_4)$ in this case reveals
that our domain wall gives rise to a D2-brane dipole moment, induced from
an effective local D2-brane charge of the wall whose value turns out to be 
\be
\label{dipole}
N_{2,{\rm eff}} =  \beta \,(4\pi\, G_{10})^{-1} \, \Big(\tau_2^{-1} \,
\pi R^2 \, V\Big).
\ee
We will see in the next subsection how this prediction for the value of 
the local D2-brane charge carried by the domain wall matches exactly with 
the microscopic properties of the constituents that sustain the domain wall.

For the second outside solution, the three-charge rotating 
extremal black hole, an equivalent calculation of the matching
condition also yields a very simple form of the effective stress-energy
tensor of the domain wall,
\be
\label{splitt}
(8\pi G_{10})\, S_{\mu\nu}^{II} = -{{\gamma_s} + {\gamma_0} +
{\gamma_4} \over R}\, \left(dt + {\beta R^2 \over 2}\sigma_3\right)^2
+2\beta R\left(dt + {\beta R^2 \over 2}\sigma_3\right)\sigma_3 
 + {\gamma_s \over R}\, dy^2 + {\gamma_4 \over R}\,
 ds^2_{T^4}.
\ee 
We see that the total stress-energy tensor of the wall
splits into a sum of individual contributions due to the three individual
charges and to the angular momentum.  This fact becomes even more transparent
in the contravariant form of the stress-energy tensor, with both indices
raised.

\subsection{The Stress-Energy of Supertubes}

Here we will compute the stress-energy tensor for a supertube
extended along the directions $t$, $y$, and wrapped on the Hopf fiber
$\phi$ (with $\sigma_3 =d\phi + \cos\theta d\psi$).
The supertube wrapping the Hopf fiber at some fixed location on the base 
$S^2$ would violate some of our mandatory $U(1)_L\times SU(2)_R$ symmetry.  
The key to restoring this symmetry is that our supertube will be 
{\it smeared\/} along the compact internal dimensions {\it and\/} along the 
$\S^2$ base of the Hopf-fibered $S^3$.  Before smearing, this supertube is 
located at a point in the base $S^2$ and in $T^4$, and carries an induced 
metric and gauge fields of the form:%
\footnote{On the worldvolume of
the D2-brane, we use the obvious choice of induced coordinates
$(\xi_0,\xi_1,\xi_2) =(t,\phi, y)$.  Note, however, that the range of the 
periodic coordinate $\phi$ is $[0,4\pi)$, which leads various factors 
of two compared to the conventional supertube worldvolume coordinates, such 
as those used in \cite{Supertube1}.}
\begin{eqnarray}
 ds_{ind}^2 &=& {\cal G}_{ab}\,d\xi^a\,d\xi^b =
 - (dt + \beta {R^2\over 2} d\phi)^2 + {R^2\over 4}\,
 d\phi^2 + dy^2 \\
 {\cal B}_2 &=& - \beta {R^2\over 2} d\phi\wedge dy,\; {\cal C}_1 =
 - \beta {R^2\over 2} d\phi, \; {\cal C}_3 =  dt \wedge dy \wedge
 \beta {R^2\over 2} d\phi \nonumber
\end{eqnarray}
The gauge invariant modified field strength on the brane is:
\begin{equation}
{\cal F} = F - {\cal B}_2 = E dt \wedge dy + {\cal B}\,dy \wedge d\phi, 
\end{equation}
where ${\cal B} = ({B\over 2} - \beta {R^2\over 2})$.  With this 
normalization, it is easy to check that the action for an individual 
supertube before smearing takes its canonical form familiar from flat 
space, including the proper overall normalization of the action.  If we 
define $\Pi$ as the canonical conjugate of $E$, this observation in particular 
implies that our quantities $\Pi,E,B$ and $R$ satisfy relations familiar from 
flat space, such as $\Pi\, B=R^2$ (assuming that the equations of 
motion are satisfied).    

The action for $N_2$ supertubes smeared over the base $S^2$ and $T^4$ is:
\begin{equation}
S = - {\cal N}\, \int e^{-\Phi} \sqrt{- {\textrm{ det}}({\cal G} +
{\cal F})}
    - {\cal N}\, \int ({\cal C}_3 + {\cal C}_1\wedge{\cal F}).
\end{equation}
where we have a included density factor which takes into account
the smearing of our solution over the volume of $S^2\times T^4$:
\begin{equation}
 {\cal N} = N_2 \tau_2\, \cdot \Big(\pi R^2 \, V\Big)^{-1}.
\end{equation}

The second term is topological and will not contribute to
the stress-energy tensor, while the first term can be re-written
as
\begin{equation}
S' = - {\cal N}\,\int \sqrt{-{\cal G} }\, \sqrt{(1 + {1\over
2}\,{\cal F}^2)}
\end{equation}
which upon variation gives
\begin{equation}
S^{\mu\nu} = \Bigg({\partial x^\mu\over \partial \xi^a}\Bigg)
             \Bigg({\partial x^\nu\over \partial \xi^b}\Bigg) \cdot
\Bigg(  {2\over\sqrt{-{\cal G}}} {\delta S' \over \delta {\cal
G}_{ab}}
           = {\cal N}\,\Big( \sqrt{(1 +  {\cal F}^2/2)}\, {\cal G}^{ab} +
             (1 + {\cal F}^2/2)^{-{1\over 2}}
              F^{ac}F^{~b}_{c} \Big )\Bigg).
\end{equation}

To compare with eq.~(\ref{ST}) we need the domain wall
stress-tensor in its covariant form, so we use the full metric $g_{\mu\nu}$
to lower its two indices.  After solving the equations of motion as in the 
probe calculation of~\cite{Supertube2}, we find that -- just as in flat 
space -- $E=1$, and $\sqrt{(1 + {\cal F}^2/2)} = {B}/R$.  These
facts allow us write the stress-energy in the following simple form, 
\begin{equation}
\label{ST2}
-{{\cal N}\over R}(B+\Pi)\left(dt+{\beta R^2 \over 2}\sigma_3\right)^2
+{\cal N}R\left(dt+{\beta R^2 \over 2}\sigma_3\right)\sigma_3+{{\cal N}\Pi 
\over R}dy^2.
\end{equation}
If we compare with the stress tensor (\ref{ST}) from the Israel
matching conditions,
\begin{equation}
-{1 \over R} (\gamma_s+\gamma_0)\left(dt+{\beta R^2 \over 2}\sigma_3\right)^2 
+2\beta R\left(dt+{\beta R^2 \over 2}\sigma_3\right)\sigma_3+
{\gamma_s \over R}dy^2,
\end{equation}
we get that
\begin{equation}
\label{match}
 (8\pi G_{10})\,{\cal N} = 2\beta,\qquad \gamma_0 = 2\beta\,
 { B},\qquad \gamma_s = 2\beta\, \Pi.
\end{equation}
Plugging this expression for $\beta$ back into the formula for the expected
local D2-brane charge of the domain wall given in Eq.~(\ref{dipole}), we get
as a nice consistency check that
\be
N_{2,{\rm eff}} = N_2.
\ee

\subsection{Adding D4-branes}

  So far we have succeeded in matching the first outside solution, the 
two-charge reduction of the BMPV black hole, with the \godel\ solution 
using the supertube described above.  The situation with a three-charge
black hole outside is more complicated.  Once we introduce
D4-brane charge, the outside solution with just a D2-brane dipole
now develops NS5 and D6 dipoles.  There is now a wide variety of
D0/D4/F1 bound states which explicitly exhibit these dipoles
using local D6/D2/NS5 charges, much as the supertube carries the
local D2-brane charge \cite{BK}.  We start with the simplest
generalization of the supertube: a dust of $T^4$-wrapped
D4-branes smeared along and co-moving with the supertube.

  In order to proceed, we will use the fact that the addition of
the D4-branes only breaks half of the supertube supersymmetry.
This means that the total stress energy of the domain wall is
controlled by the BPS algebra: it is just the sum of (\ref{ST2})
with that of plain vanilla D4-branes smeared along the $y$-direction
and the three-sphere of radius $R$. The stress energy for the
D4-branes is 
\be
\label{ST3}
n_4\,(\tau_4^{-1}\cdot2\pi^2 R^3\cdot 2\pi R_9 )^{-1} \left\{
-\left(dt+{\beta R^2 \over 2}\sigma_3\right)^2+\sum_{i=5}^8(dx^i)^2\right\}
\ee 
The total stress-energy will match if in addition to
eq.~(\ref{match}), we have
\begin{equation}
 n_4 \,=\, \gamma_4\,{g}_s\,  {8\pi^6 (\alpha')^{5/2} {R}^2 {R}_9
 \over {G}_{10}} = \gamma_4 R^2 {R^9\over g_s l_s^3},
\end{equation}
which is exactly what we have in eq.~(\ref{n4}), as expected from
the BPS condition that mass equals charge. Thus, the microscopic details
of the bound state are largely irrelevant for the macroscopic behavior of
the domain wall.  For any general domain wall with the same amount of
supersymmetry, made up of any collection of D0/F1/D4 bound states, the
stress-energy always splits into the form,
\begin{equation}
S^{\mu\nu} = S^{\mu\nu}_0 + S^{\mu\nu}_s + S^{\mu\nu}_j +
S^{\mu\nu}_4
\end{equation}
manifested already in \ Eq.~(\ref{splitt}).
The particular microscopic characteristics of the bound will only determine
a macroscopic relation between the angular momentum and other macroscopic
parameters of the configuration.  Only this macroscopic condition is needed
to ensure the validity of the Israel matching conditions.

\section{Microstates and the Asymptotic Bulk Fields}

We now want to understand our domain wall matching in the context of supertube
microstates.  More generally, we aim at a deeper understanding of
the relationship between various detailed microstates which come
from bound states of F1,D0 and D4-branes, and the asymptotic bulk
fields which they source.  The first step is to count
configurations with large $U(1)_L$ angular momentum $j_L$.
We can facilitate this process by counting U-dual states with the same 
angular momentum.

We first examine the case when all the angular momentum of a
putative three-charge black hole ground state is carried by just one
type of charged component.  For example, suppose we U-dualize such
that all the angular momentum is carried by wrapped F1-strings. In
order to transform under the same super-conformal algebra as the 
black hole, these must be in the RR ground state.  These states
transform in the ({\bf 1},{\bf 1}), ({\bf 3},{\bf 1}), ({\bf
1},{\bf 3}) and ({\bf 2},{\bf 2}) of $SU(2)_L\times SU(2)_R$. This
implies that the maximum $U(1)_L$ charge carried by $n_i$
microstate components alone is
 \be
 \label{onecharge}
 j_L = 2 n_i,\;\textrm{with}\quad j_R = 0.
 \ee

    Next we consider cases where angular momentum is carried by
bound states of two charges $(n_i,n_j)$.  We take as our canonical
example the D1/KK(momentum) system ($n_i = n_1, n_j =n_{KK}$).
One important characteristic of this system comes from the zero
modes of the D1-D1 open strings. To get the maximal angular
momentum, we wind a single D1 string $n_1$ times, and then use the
fact that the mimimum KK-momentum for the left-moving strings is
now $1/n_1R$ to carry this KK-momentum with $n_1n_{KK}$ open
strings. Each of these has angular momentum in the ({\bf
1},{\bf 1}) or $({\bf 2},{\bf 2})$ if it is bosonic or in the
({\bf 1},{\bf 2}) or ({\bf 2},{\bf 1}) if it is fermionic. This
means that the maximum value we can achieve for $j_L$ is now
 \be
 \label{twocharges}
 j_L = n_1 n_{KK} = n_in_j.
 \ee
For such a state, bosonic and fermionic statistics imply that
$j_R$ must take the value $n_in_j$, $n_in_j -1$, or $n_in_j - 2$. With
two fermionic bound states, we can make $j_R$ very close to zero.
In the maximal angular momentum case above, we should point out
that the U-dual D1/D5 system maximally fragments into $n_in_j$
string bits, each carrying the maximum amount of angular momentum
(for more details see~\cite{Mathur}).

    Finally, we consider the angular momentum carried by three
bound charges with four supersymmetries.  Our canonical
choice is now the D1/D5/KK system excited so as to carry as much
angular momentum as possible. In this case, one long string in the
RR sector\cite{MS}, with length $2\pi R_9n_1n_5n_{k}$, carries
all the angular momentum through left-handed fermionic strings
with one unit of the $U(1)_L$ charge.  The right handed sector has $U(1)_R$
charge $\pm 1$. In the limit where one of the charges, such as
$n_{k}$ is much larger than the product of the other two,
the maximum angular momentum is \cite{BMPV1}
\be
\label{threecharges}
j_l = 2\sqrt{n_in_jn_k},
\ee
with $j_R = \pm 1$.

  We finish this summary by remarking on the effectiveness of
various bound states at carrying angular momentum.  If we start by
carrying all the angular momentum with two charges, we can compare
eqs. (\ref{onecharge}) and (\ref{twocharges}):
 \be
   n_i n_j > 2 n_i + 2 n_j\;\;\textrm{for}\quad n_i,n_j > 3.
 \ee
This inequality means that different sets of multiple charges can
almost always carry more angular momentum by binding together than
they could alone.  When we move to three charges, this binding
advantage no longer seems to apply in all cases.  For example, if
we take one charge much larger than the others
\be
  n_3 \gg {n_1n_2}
\ee
then we can increase the overall $j_L$ by placing all of the
angular momentum in the third charge since
\be
  2 n_3 = 2\, (\sqrt{n_3})^2 \gg 2 \sqrt{n_1n_2n_3}.
\ee

\subsection{Interpretation of Matching Part I: The Two-Charge Domain Wall}

With the context of various black hole microstates now in mind,
we can now interpret the results of our matching for the first
outside solution, the two-charge reduction of the rotating black hole.  
Apart from the proper matching of the D0-brane and F1-string charges we get
two important results which relate the microscopics of the domain
wall with the parameters of the bulk solution.

The first result comes from the internal dynamics of the supertube,
$R^2 = \Pi B$, plus the matching of D0 and F1 charges.  It yields
 \be
 \label{result1}
 \gamma_s  \, \gamma_0 = 4\beta^2 R^2
 \ee
We note immediately that by definition we have $\gamma_0,\gamma_s
< 1$.   This inequality combined with the equation above directly
implies that $R < 1/2\beta$.  Hence, we never have closed timelike
curves inside the domain wall (and thus no CTCs outside either)!
The redshifting of central charges, $\gamma_i = Q_i/(Q_i +
\tilde{R}^2)$, allows us to extract one more piece of information
out of eq.~(\ref{result1}): the relationship between the principal
macroscopic dimensionful quantities,
\begin{equation}
\label{macrocharges}
 Q_s\,Q_0\,\tilde{R}^2 = J^2.
\end{equation}
Comparing with eq.~(\ref{rc}), we can check that $\tilde{R}$ is
clearly larger than the critical radius $r_c$.  Here, the squared
radius of the domain wall (related to the D2 dipole, as we will see)
plays much the same role in the macroscopic charge relations as
the D4-brane charge does for the three-charge BMPV black hole.   In
fact, in the next subsection when we add in spectator D4-brane
charge, the domain wall radius squared and the D4-brane charge
will appear additively.

The second result from our matching data, eq.~(\ref{match}), is
that:
\begin{equation}
\label{betavalue}
\beta R^2 = \,{4 N_2 G_6 \tau_2 },
\end{equation}
If we combine this relation with eq.~(\ref{result1}) we get two
separate equations for the integral value of the angular momentum.
The first one,
 \be \label{j1}
j = {\pi J\over 4 \tilde{G}_{5}} = {\pi \over 2 \tilde{G}_{5}}
{\beta} \tilde{R}^4 \chi_s^{1/2} \chi_0^{3/4} = n_0\,n_s/N_2 =
N_2\, \Big({n_0\over N_2}\Big)\Big({n_s\over N_2}\Big) \;,
 \ee
is exactly like the expression for a supertube with rank-two
angular momentum in flat space. According to eq.~(\ref{twocharges}), this
is also the maximum angular momentum allowed for any microstate which comes
from $N_2$ bound-states of $n_0/N_2$ D0-branes and $n_s/N_2$ fundamental 
strings.  The second equation is
\begin{equation}
\label{j2}
  j = {\pi J\over 4 \tilde{G}_{5}} = {\pi^2 J \tilde{R}_9\over 2 \tilde{G}_6}
    = N_2 \,\tau_2 \, 4\pi^2 {R}_9\,{R}^2,
\end{equation}
which shows that $j$ is the sum of angular momenta from the  $N_2$
$D2$-branes with identical dipole moments.

To recap, the classical BMPV black hole solution with only two charges
$Q_0, Q_s$ and nonzero angular momentum $J$ has a naked singularity at $r=0$ 
and exhibits closed time-like curves in the vicinity of the singularity (and
consequently throughout the entire spacetime).  For a sufficiently small, yet
macroscopic angular momentum we find that the classical 
solution can be naturally truncated at a fixed radius by a smeared supertube,
removing all closed time-like curves. An interesting thing to note
here is that the microstate condition $j N \le n_0 n_s$, gives a
bound on the asymptotic charges $J,Q_0,Q_s$ which depends on
the moduli $\tilde{g}_s$ and $\tilde{V}$:
\begin{equation}
J \le  {Q_0 Q_s\over 8 \tilde{\tau}_2\tilde{G}_6}.
\end{equation}

Of course, here we have only described the resolution of the BMPV solution 
for the case $N j_L= n_0n_s$.  For smaller amounts of angular momentum, 
a circular supertube carrying all the charge will not work.  To remedy this, 
we can either increase the amount of charge free (unbound) from the 
supertube, or change the shape of the supertube (which reduces its angular 
momentum \cite{Supertube3,Supercurves,PM}). D0-branes and F1-strings are 
BPS states with compatible supersymmetries, so they can be placed anywhere and
they will feel no force.  For sufficiently large numbers of these
free charges, this means we could form supersymmetric two-charge
\godel\ black holes inside our domain wall. On the other hand, we
can also make ripples on the D2-brane supertube which will lower
its angular momentum but asymptotically only contribute to higher
D2-brane moments. Both configurations will quickly start to look
like the appropriate BMPV black hole asymptotically. In order to
preserve an exact $SU(2)_R$, all these configurations
still need to be smeared.  Only the un-smeared rippled supertubes
are proper boundstate microstates with no loose charges;
they are U-dual to the black hole hair of Mathur and
Lunin~\cite{ML1,ML2,ML3,ML4,ML5,ML6}.

We end our discussion of the two-charge case by pointing out that in
addition to our smearing, the asymptotic solution is further
coarse-grained by the fact that it only relays classical
information about angular momentum. An observer far enough away
will not be able to distinguish between our domain wall and one
with $m_R \ll n_0 n_s$.  It is possible that the 3-sphere area of
our smeared supertube is related to the number of D0/F1 bound
states which are identical under this further coarse graining.

\subsection{Interpretation of Matching Part II: The Three Charge Domain Wall}

    We now proceed to consider the domain wall match between
the three-charge BMPV black hole and our five-dimensional \godel\
universe.  Contrasted with the the two-charge case, there is now a
greater variety of D0/F1/D4 bound state micro-states which we
can smear to preserve an exact $U(1)_L\times SU(2)_R$ R-symmetry.
This implies a whole zoo of possible objects uniformly distributed
on a squashed three-sphere which we could use as a domain wall.
We have given the exact details for only one case, a D0-F1
supertube with an unbound dust of $T^4$-wrapped D4-branes. In
this subsection, we will show how the details of this
specific choice of microstate feed backs into the parameters of
the bulk solution.  We will explore what a more general domain
wall implies for these parameters in the next subsection.

  We have chosen to put our domain wall outside or on the
horizon, so we again have $\gamma_0,\gamma_s,\gamma_4 \le 1$. Since
we have not modified the supertube, and are dealing with BPS
states, the matching conditions yield exactly the same form as
before for the parameter $\beta$, and we still have an ``equation
of state'' for the supertube of the form $4\beta^2 R^2 =
\gamma_0\gamma_s$. However, the redshift factors are now
different: this changes the equation relating the macroscopic
dimensionful quantities in the asymptotic coordinate system.
Indeed, now
\begin{equation}
\label{macro2}
J^2 = Q_s Q_0 \tilde{R}^2 \chi_4^{-1} = \gamma_4^{-1} \cdot Q_sQ_0Q_4 =
{Q_sQ_0\,(Q_4 + \tilde{R}^2)}.
\end{equation}
Once again, the domain wall radius is such that no closed
time-like curves appear.  For a more general three charge
supertube, this equation of state will have to be modified to reflect
the details of the bound state which carries the angular momentum.

    The second item we get from matching at the domain wall is the
value of $\beta R^2$ in terms of the moduli at the domain wall.
This just comes from matching the stress-energy tensor: for our
choice of the domain wall (smearing $N_2$ supertubes plus $D4$-dust)
we have
\begin{equation}
\beta R^2 = \,{4 N_2\, G_6\, \tau_2 }.
\end{equation}
This give the following two equations for $j$
\begin{equation}
n_0n_s/N_2 = j  = N_2 \, {4\pi^2\tilde{R}_9\, \tilde{\tau}_2 } \,
\tilde{R}^2  = N_2 \,\tau_2  \, 4\pi^2 {R}_9\, {R}^2,
\end{equation}
reflecting that fact that all the angular momentum is being
carried by the D0/F1 bound-state D2-dipoles.  The D4-branes
carry no extra angular momentum.

  Going back to eq.~(\ref{macro2}), we see that for a dipole of
sufficiently large radius, the three-charge BMPV black hole will be
over-rotating.  Once again, we find that our domain wall shell
acts as a guardian of chronology, cutting off the chronology-violating
region of the classical solution, and replacing it with a causal core.

It may also be instructive to consider under which conditions the
black hole will be under-rotating.  (For simplicity's sake, we will
now assume $N_2 =1$.)  In order that the total angular momentum is
small enough to guarantee the existence of a horizon, eq.
(\ref{macro2}) shows that we must have:
\be
\gamma_4 > 1/4.
\ee
In terms of integral charges, this implies
\be
n_4 > {n_0n_s}/4\, .
\ee
For such large $n_4$ the bound on the angular momentum for a
bound-state of $n_0$ D0-branes and $n_s$ F1-strings with $n_4$
wrapped D4-branes is
\be
j \le  2 \sqrt{n_0 n_s n_4}
\ee
so it is possible for this bound-state to carry at least as much
momentum as a supertube of bound D0's and F1's alone.

These conditions on over-rotation vs.\ under-rotation are
consistent with our microscopic understanding of black holes for
the following reason. If we were to perturb our shell with a
little bit of extra energy, two things could happen.  If $n_4$ is
large enough, the D4-branes can bind with the branes in the
supertube and the whole object becomes a microstate of an
under-rotating black hole since these are D4/D0/F1 bound-states.
On the other hand, if $n_4$ is too small, any black hole we try to
form will leave a shell of rotating matter behind.  The asymptotic
fields will be that of an over-rotating black hole, but
chronology will still be saved by a domain wall.  These will be
solutions corresponding to \godel\ black
holes~\cite{Herdeiro,GH,GodelBH2,BDGO,BD} separated from an
over-rotating BMPV black hole by a domain wall.

\subsection{Other Domain Walls}

  In \cite{Dyson}, another possible mechanism was offered for
preventing the creation of over-rotating black hole.  When the author
of \cite{Dyson} tried to bring in extra angular and KK-momentum on
adiabatically collapsing shell, she discovered an
enhan\c{c}on-like effect which encouraged the shell to remain at a
radius\footnote{We propose that the resulting sphere should be called a
Dyson sphere.}
which, after a T-duality transformation mapping
D1/D5/KK to D0/D4/F1, solves the equation
\begin{equation}
J^2 = 4 Q_s (Q_0 + \tilde{R}^2_D)(Q_4 + \tilde{R}^2_D)
\end{equation}
In this instance, the bound on the angular momentum is once again
circumvented by sequestering proposed extra mass and angular
momentum in a shell outside the horizon.  Even if we place all the
charge on this shell, the solution inside is no longer a \godel\
solution, it is a homogenous fluxbrane solution (a different
T-duality takes this to flat space).

In the context of our work, we see that this time the angular
momentum is carried by a single charge $n_s$.  The maximum
groundstate $U(1)_L$ spin carried by these charges happens when
$j_L = 2n_s, j_R=0$. A little hard work gives us equations very
similar to those from a supertube domain wall:
\be
  \beta^2 R^2 = \gamma_s,\qquad  2n_s = j = {\pi R_9^2\over \alpha' G_6}\,
  R^4.
  \ee
Note that now $j$ grows with the fourth power of $R$.  A U-duality
transformation gives us similar monopole shell resolution for
D0's and D4's.  In those cases, the angular momentum comes
from the ground-state zero modes

  As opposed to the supertube dipoles, the enhan\c{c}ons
or Dyson spheres only appear with positive $\tilde{R}^2$
for over-rotating three-charge BMPV black holes.
Depending on the size of the single charge in
question, though, they will truncate the solution either inside or
outside the radius at which our supertubes truncate it.  It would
be interesting to understand the physical meaning of the corresponding
cross-over point.

   Apart from these individually spinning-charge domain walls, there are
of course other dipole domain walls which come from U-duality.
For example, a simple T-duality transformation takes the D2-dipole
which is a bound-state of D0's and F1's to a D6-dipole which is a
bound-state of D4's and F1's \cite{BK}.  All the possible dipole
domain walls will have a similar \godel\ solution inside. A quick
way to understand this is to look at the \godel\ solution with the
choice of RR-fields in eq.~(\ref{10Dmetric}).  If we lift this
solution to M-theory, there is $U(3)$ symmetry which mixes the
three two-planes $x^6x^7, x^8x^9$ and $x^5x^{11}$.  The discrete
subgroup of this symmetry which exchanges the two-planes becomes
in our main solutions the U-duality which exchanges D0, D4 and F1
charges; this maps \godel\ back to itself modulo some possible
signs in the various forms.

  Finally, there is the possibility of a domain wall which is the
maximally spinning bound state of three charges.  We call this
microstate the hypertube. Presumably this state should correspond
to a D0/D4/F1 bound-state discussed before.  If we pick $n_0 \gg
n_4 n_s$, we know that the state which saturates the angular
momentum bound has $j^2 = 4 n_0 n_s n_4$.  Unfortunately, in the
right-moving sector this state will always carry some angular
momentum in $U(1)_R$ and in the $x^{6..9}$ four-plane, and thus
does not represent a perfect candidate for a domain wall which
preserves an exact $U(1)_L\times SU(2)_R$ symmetry. We can,
though, take two of these hypertubes and then pick a quantum state
with all the angular momentum from the right-moving sector set to
zero.  We can also just smear a single hypertube.

    The details of a hypertube domain-wall as realized on a
D-brane are not yet known, and beyond the scope of this paper,
but by clever inference we can draw some conclusions as to the
properties of a collection of any number, $N_H$, of hypertubes.
First of all, their angular momentum is bounded by:
\be
j \le  2 N_H\,\sqrt{{n_0\over N_H}{n_s\over N_H}{n_4\over N_H}} =
2\sqrt{ n_0n_sn_4\over N_H}.
\ee
which using our definitions implies
\be
R \le \sqrt{\gamma_0\gamma_s\gamma_4/N_H}\,{1\over
2\beta}\qquad\textrm{and}\qquad J^2 \le 4 Q_0Q_sQ_4/N_H.
\ee
A domain wall with $N_H$ hypertubes always cuts off an
under-rotating black hole, consistent with our understanding that
with a little extra energy the $N_H$ parts of the domain-wall
could form a single bound state which would retreat behind the
horizon of this putative black hole.  The single hypertube exactly
saturates the rotation bound, where the black hole horizon area
disappears, leaving a diminished number of possible microstates.  We
also see that hypertube domain walls outside the horizon always
bound a \godel\ universe with radius $R \le 1/2\beta$.

\section{The Decoupling Limit}
\label{limit}

  In this section, we will look at how our domain wall matching
plays out in the near-horizon limit of the BMPV black holes we
consider.  For the two-charge solution, there is really no
horizon, since the dilaton blows up as we take $\tilde{r}\to 0$.
On the other hand, the U-dual D1/D5 system is a solution with a
nice decoupling limit \cite{Maldacena}.  It is instructive to see
what this U-dual limit does in our case.  For the three-charge BMPV
black hole, the limit is simpler, and yields a spacetime which is
a deformation of $AdS_2\times S^3\times T^5$.

\subsection{The U-duality Map and Decoupling Limit of the
two-Charge Solution}

  The two-charge solution we wrote down in 
eq.~(\ref{BMPV}) has a singularity at the origin, cut-off by our
domain wall of course.  One might understandably be shy of
performing a "near-singularity" limit.  Still, we will argue that
we can take such limit smoothly by taking the $\gamma_i \to 1$. To
help motivate this choice, we look at our solution through the
looking-glass of the U-dual D1/D5 system, and infer the
parameter and coordinate scalings of our solution from those of
the Maldacena limit of its U-dual.

A chain of dualities which maps the the D0/F1 wrapping the
$y$($=x_9$) direction system to the D5/D1 wrapping
$\bar{x}_5,\ldots\bar{x}_9$ and $\bar{x}_5$ is as follows. First we
T-dualize along the 678-directions to get D3/F1, then S-dualize to
get D3/D1 and finally a T-duality along the 59-directions gets us
to D5/D1. We denote the original $T^4$ radii $\tilde{L}$.  The
D1/D5 compact directions have $\bar{R}_5 = T, \bar{R}_9 = S$ and
$\bar{R}_{6...8} = K$.  All these dualities are applied in the
string frame, so the S-duality will introduce a non-trivial scale
transformation involving the ratio:
\begin{equation}
\rho = \sqrt{{S T\over \bar{g}_s l_s^2}} = \sqrt{ \tilde{g}_s
l_s^3\over \tilde{L}^3}.
\end{equation}
Using this ratio, we can now write the parameters and coordinates
of the D1/D5 solution in terms of those of our D0/F1 charged black hole
\begin{equation}
\label{udual1}
 S= \rho {l_s^2\over \tilde{R}_9}, \quad T= \rho
{l_s^2\over \tilde{R}_5 },\quad K = \rho^{-1} {l_s^2\over
\tilde{L}}, \quad \bar{g}_s =  {l_s^2 \over \tilde{R}_5\tilde{R}_9
},
\end{equation}
and
\be
\label{udual2}
\bar{x}_{0...4} = \rho^{-1}\, \tilde{x}_{0...4},\qquad \bar{Q}_i =
\rho^{-2} {Q}_i.
\ee
We will now show that the Maldacena limit corresponds after
U-duality to a scaling of our solution which takes $\gamma_s,
\gamma_0 \to 1$ while keeping fixed the \godel\ geometry and
couplings in the near-tube coordinates.  We call this the
decoupling limit.

The Maldacena limit scaling can be written as taking $\e \to 0$
with
\begin{equation}
\label{mald}
\bar{r} \to \e \bar{r},\quad \bar{t} \to \e^{-1} \bar{t},\quad \bar{T} \to
\e^{-1}\bar{T},\quad \bar{K},\bar{S} \to \e^{0} \bar{K}, \bar{S}
\end{equation}
We can model this scaling for our coordinates as follows.   The
first step is to write both $\gamma_0$ and $\gamma_s$ as $1 -
\e^2$, then we write the asymptotic coordinates in terms of the
near-tube ones as:
 \begin{eqnarray}
 &&
 Q_i = \e^{-1} R^2,\; \tilde{r} = \e^{1/2} r,\quad\tilde{t} = \e^{-3/2}\, t,\quad \rho = \e^{-1/2}
 \sqrt{g_sl_s^3/L^3} \nonumber \\
 && \tilde{L} = \e^{1/2} L, \quad\tilde{R_5} = \e^{1/2} R_5, \quad\tilde{R}_9 =
 \e^{-1/2} R_9,
 \end{eqnarray}
Plugging these values into Eqs.~(\ref{udual1}) and (\ref{udual2}),
we see that we get
\begin{eqnarray}
 && \bar{Q}_i = {L^3\over g_s l_s^3} R^2,\quad \bar{r} = \e
 \sqrt{L^3\over g_s l_s^3} r, \quad\bar{t} = \e^{-1} \sqrt{L^3\over g_s
 l_s^3}t, \\
 && \bar{T} = \e^{-1} \sqrt{g_s l_s^7\over R_5^2 L^3},\quad
    \bar{S} = \sqrt{g_s l_s^7\over R_9^2 L^3}, \quad
    \bar{K} = \sqrt{g_s l_s^7\over L^5},
     \nonumber
\end{eqnarray}
just as required to satisfy Eq.~(\ref{mald}). 

After we take the $\gamma_i \to 1$ limit, or near-horizon limit,
we get new metric and potentials which (after some gauge
transformations) look like
\begin{eqnarray}
ds^2 &=& - {r^3\over R^3}\, (dt +  {R^3\over 4 r^2}\, \sigma_3
)^2+ {r \over R}\, dy^2
+ {R\over r} \sum_{i = 1}^8 (dx^i)^2, \\
B_2 &=& - {r^2\over R^2}\,(dt + {R^3\over 4 r^2}\, \sigma_3)\wedge
dy  + dt\wedge dy
 ,\qquad e^{\Phi} = g_s\,\sqrt{{R\over r}}, \nonumber\\
C_1 &=& - {r^2\over R^2}\, (dt +  {R^3\over 4 r^2}\, \sigma_3)
 + dt
,\qquad C_3 = {R\over 4}\, dt\wedge dy \wedge \sigma_3
\nonumber.
\end{eqnarray}
This solution still has a singularity at $r=0$, but it no longer has
any closed time-like curves.  The singularity will be removed by our
domain wall at $r = R$. From eq.~(\ref{betavalue}), we see that
\be
R = 8 N_2 G_6 \tau_2^{-1}
\ee
and so can be made arbitrarily large by increasing the number of
D2-branes.

In the U-dual D1/D5 background, the asymptotic geometry is locally
$AdS_3\times S^3\times T^4$, and our D2-brane shell now becomes a
KK-monopole~\cite{ML6}.  Since the KK-monopole is a supergravity object,
when $N_2=1$ that background is actually smooth, and the domain wall
disappears.

\subsection{The Three-Charge Near Horizon Limit}

 Now that we have identified the right procedure for the
decoupling limit in the two-charge solution, it is simple to
extend this procedure to the three charge black hole.  We just switch to
the near-tube coordinates and then take the limit $\gamma_i \to
1$. This can be checked using the U-duality above, or more simply,
by T-dualizing along the $y$ direction.  For our choice of domain
walls, this limit gives:
\be
U,V,W \to  {R^2\over r^2},\qquad \beta R \to {1\over 2}.
\end{equation}
Then the metric and fields (modulo gauge transformations) for the
outside solution takes the simple form
\begin{eqnarray}
 ds^2 &=& - {r^4\over R^4}
 (dt +  {R^3\over 4 r^2}\, \sigma_3)^2+ dy^2
 + R^2 {dr^2\over r^2} + {R^2\over 4}(\sigma_1^2 + \sigma_2^2 +
 \sigma_3^2)
 + \sum_{i = 5}^8 (d\tilde{x}^i)^2, \\
 B_2 &=& -{r^2\over R^2}(dt +  {R^3\over 4 r^2}\,\sigma_3)\wedge dy
  + dt\wedge dy
 , \qquad e^{\Phi} = g_s, \nonumber \\
 C_1 &=& - {r^2\over R^2} (dt +  {R^3\over 4 r^2}\,\sigma_3)
 + dt
 , \quad C_3 = {R^2 \over 4}\, \cos\theta\, d\phi\wedge d\psi\wedge dy
  + {R\over 4}\, dt\wedge dy \wedge \sigma_3
 \nonumber
\end{eqnarray}
which is a deformation of $AdS_2\times S^3 \times T^5$.  We expect
this space-time to be dual to some deformation of a
superconformal quantum mechanics.  It would be interesting to
understand the appearance of a domain wall at $r=R$ in the
supergravity solution in terms of that quantum mechanics.
Alternatively, a T-duality takes this background to a geometry
which is locally $AdS_3\times S^3\times T^4$.  The appearance of
our domain wall now corresponds to the nucleation of long
D-strings very similar to those in~\cite{Israel}.


\section{Supertube Quantum Mechanics, Hypertubes and \godel\ Holography}

   In order to better understand how our domain wall solutions relate to
micro-states of BMPV black holes and AdS geometries, a careful examination 
of these domain walls as states in the D2-brane DBI theory is in order.  
The explicit states we have seen so far are smeared D2-branes; we will 
contemplate this smearing in more detail.  

   A circular supertube is a compact extended object.  As such, it has not 
just a center-of-mass location, but also a size, orientation, and a 
world-volume gauge field.  It's world-volume fields include four periodic 
scalars whose vevs determine its location on $T^4$;  we smear by averaging
over this location.  In $R^4$, the supertube is a circular ring specified by 
a position vector, a radius $R$ and an orientation which takes values in 
$S^2$.  All of these are scalars of the D2-brane worldvolume theory.  The 
position decouples from the dynamics which interest us, therefore 
we will ignore it.  As demonstrated in~\cite{Supertube1,Supertube2}, 
the radius field $R$ has a non-zero vev due to a potential which depends on 
the charges $\Pi$ and $B$.  The circle which the supertube wraps is 
non-trivially fibered over the orientation $S^2$: this is actually the Hopf 
fibration with total space $S^3$.  This implies that angular momentum in one 
two-plane is parallel transported into other two-planes as the tube moves
in $S^2$.  Smearing our supertube source over the orientation $S^2$ then 
gives it angular momentum in $U(1)_L$ and preserves the $SU(2)_R$ symmetry of 
our supergravity solutions.

A single smeared supertube can carry $SO(4)$ angular momentum with
rank 4, while a localized supertube is restricted to rank 2  
angular momenta.  In order to accomplish our smearing over $S^2$
we would have to use some quantum mechanical averaging whose exact nature 
is not immediately clear\footnote{The corresponding state may not
even be a pure state}. More conventionally, a collection of a large number of 
$N_2$ separate supertubes, distributed around the $S^2$ and smeared along the 
flat directions, can also give a 
source with rank 4 angular momentum~\cite{BK}. When $N_2$ is
sufficiently large, we can spread the supertubes out on $T^4$ and
the orientation $S^2$, obtaining a good classical approximation of
$N_2$ smeared supertubes. Since rings with different orientations
in $R^4$ preserve the same supersymmetries, the rings in our
collection feel no relative force and retain the same energy as
$N_2$ smeared supertube.  The advantage of using a collection of
$N_2$ un-smeared supertubes is that we can trust the supergravity 
solution a lot closer to the domain wall than when we try to pile up 
$N_2$ individually smeared 
supertubes.  In the latter case, we should not trust our solution
any closer than the size of region we smear over, which grows like
$R$.  In the former case, we can trust our solution up to the
spacing, $\ell$, separating any two supertubes, which is of order
\be
\ell = \Big({\pi R^2 V\over N_2}\Big)^{1/6}.
\ee

  For hypertubes, less smearing is required.  Whereas a supertube with a
dust of co-moving $T^4$-wrapped D4-branes has angular momenta
$j_L$ and $j_R$ both of the same order, and thus tends to extend
in only one two-plane, once the tube binds with the D4-branes to
form a supersymmetric hypertube we can only have one large angular
momentum, either $j_L$ or $j_R$.  For $N_H$ hypertubes we have
\be
j_L \le 2\sqrt{n_0n_sn_4}/N_H^{1/2},\qquad j_R \le N_H
\ee
This means that for $n_i \gg N_H$, hypertubes have mostly rank 4
angular momenta and will extend in all the $R^4$ directions.
Hypertubes are necessarily domain walls!  This complements well
with the U-dual results of~\cite{Lunin}. 

\subsection{Limits of Validity: Theory on Supertube}

Our construction provides a mechanism for resolving the problem of closed 
time-like curves in the \godel\ universe and in over-rotating BMPV black 
holes.  In fact, we see that the two classical solutions solve each other's 
causality problems.  
This involves the appearance of smeared supertube domain wall between 
causally consistent portions of the two solutions, and
possibly even more general objects.  A natural question is that of the 
limits of validity, apart from those due to smearing, for such supergravity 
domain walls.  Domain walls in general
present a very interesting case for the duality between D-brane
world-volume theories and the supergravity backgrounds that they
source.  As we increase the closed string coupling for D-branes
with codimension larger then one, the surrounding background
starts to curve, and a throat region emerges.  For domain walls,
the full gravitational backreaction is taken care of by solving
the Israel matching conditions: no throat region appears.  We can
no longer think about replacing the D-brane with a dual
near-horizon background, the D-brane and the curved background
coexist. Stringy corrections can become important, of course, when
the energy density of the domain wall reaches the string scale.

  Our domain walls are in a sense an interesting hybrid because they are
compact along three worldvolume dimensions.  Far away, they look
like bound states of D-branes which (after wrapping/smearing in
the internal dimensions) appear approximately pointlike in the directions 
transverse to the noncompact dimensions of the domain wall.  Hence, the 
backreaction on the geometry makes their asymptotic fields identical to that 
of a black hole (or a continuation thereof, to the regime in which the 
naive classical solution would exhibit a naked singularity surrounded by 
naked closed timelike curves).  Once we get close to the domain wall, the
back-reaction is instead encoded in the jump of the extensive curvature,
which for our solutions is of order $1/R$.  As long as $R$ is
large in string units, we don't need any stringy corrections and
the thin wall approximation should hold.  In our case, we can
always make $R$ sufficiently large, and the smearing less
egregious, by increasing $N_2$, the number of overlapping
supertubes\footnote{For supertubes, the large $N_2$ limit corresponds to
the long string limit of the U-dual D1/D5 system}.
On the other hand, for $N_2 = 1 $, we have in the near-tube limit
 \be
   R/l_s \propto g_s (l_s^4/V)
 \ee
which will always be small in the perturbative regime $g_s \ll 1 $.  
If we try to make $V \ll l_s^4$, we are forced to T-dualize and the problem 
remains.  At this point, the region near the origin is inherently stringy, 
and is best described in terms of the non-commutative quantum mechanics of 
$n_0$ D0-branes~\cite{BL}.

We can further corroborate our view of the limits of validity if we consider 
the validity of the open string expansion for our domain wall.  Since there 
is a non-trivial background B-field and magnetic field on the worldvolume 
of our supertubes, it is best to consider the expansion in the variables
of Seiberg and Witten~\cite{SWnc,Aspects}.  Using these variables, we see 
that open string coupling for a single localized supertube is:
 \be
  g_s \Bigg({\det({\cal G} + 2\pi\alpha'{\cal F})\over 
\det({\cal G})}\Bigg)^{1/2}  = g_s \, {B\over R}.
 \ee
If we take into account the fact that we have have $N_2$
supertubes, as well as the fact that these are spread%
\footnote{When we smear or spread D-branes, the number of choices
for an open string to end on is effectively the density of
D-branes in string units.}
over a volume $\pi R^2V$, we see that this open string coupling needs to be
adjusted by a multiplicative factor which takes the form (up to minor 
numerical factors):
 \be
    N_2\,{(2\pi\alpha')^3\over \pi R^2 V} = \pi^{-2}g_s^{-1}\beta l_s.
 \ee
This means that the open string coupling constant for the smeared supertubes 
is approximately
\be
  G_o = {\gamma_0\over 2\pi^2R} l_s,
\ee
which is of the same order as the jump in the extrinsic curvature of the 
domain wall in string units.  Therefore, we see that our whole picture fits 
together quite well for small $g_s$ and large $R$.  For small $N_2$, we once 
again have to go to the quantum-mechanical description in terms of D0-branes.

\subsection{\godel\ Holography}

Now that we have a better understanding of what kind of domain walls we 
should use to cut-off a \godel\ solution, we could speculate on what this 
means for applications holography to these solutions.  By finding our domain 
walls, we have accomplished the first step in our systematic program to 
understand the quantum states of a stringy \godel\ universe.
Note that our construction allows one to make sense of \godel\ solutions 
which are of {\it arbitrarily large\/} size compared to the string scale, 
simply by dialing $N_2$.  We have also identified a near-wall limit which 
should prove useful in the process of isolating the stringy quantum states of 
one \godel\ region of radius $1/2\beta$.  For large $N_2$ and small $\beta$,  
we can characterize the degrees of freedom of the domain wall in terms of a 
non-commutative 2+1 field theory, while for small $N_2$ and large 
$\beta$, the description should be in terms of a D0-brane matrix 
quantum mechanics instead.   One should be able to obtain and study the 
partition function exact for any $N$ by generalizing the boundary states 
of~\cite{Takayanagi}.

If our domain walls indeed represent natural holographic screens for the 
\godel\ universe, one interesting fact about them is that they are located 
at radius $1/2\beta$, instead of the value $\sqrt{3}/2\beta$ expected from 
the ``phenomenological'' analysis of preferred holographic screens as 
predicted by the behavior of massless, point-like probes of the classical 
\godel\ geometry \cite{BGHV}.  As we remarked on above, stringy probes 
certainly see the \godel\ geometry somewhat differently than point-like 
probes, and a possible finite renormalization of the location of the 
holographic screens should not be unexpected.   Since
our domain wall is made up of smeared objects wound along orbits
with line elements proportional to $\sigma_3$, it perhaps not so
surprising that our screens appear before the radius grows to the value 
predicted by the point-like probes.  

\section{Final Thoughts}

We have argued in this paper that supertubes, and their more general cousins, 
provide a microscopic string-theory resolution of the causality problems 
found in a large class of supersymmetric classical solutions.  In particular, 
we have demonstrated that the \godel\ universe and the class of 
over-rotating BMPV black holes solve each other's causality problems, by 
joining along a supertube domain wall.  

This mechanism has interesting implications both for the outside black-hole 
solution, and for the inside \godel\ geometry.  From the outside perspective, 
it provides a string theory resolution to a broad class of supersymmetric 
timelike naked singularities describing over-rotating black holes, which would 
be otherwise thrown away in classical general relativity by the assumption of 
cosmic censorship.  

The implications for the \godel\ solution are perhaps even 
more profound.  The classical \godel\ solution is homogeneous in spacetime, 
and thus unique for a fixed value of its vorticity $\beta$, but suffers from 
closed timelike curves and therefore its consistency is in question.  Based on 
the resolution mechanism presented in this paper, we propose that in full 
string theory the unique (but inconsistent) classical solution is replaced by 
an entire moduli space of causally consistent domain-wall resolutions.  The moduli are 
given by the location and geometry of the domain wall that connects the causal 
part of the \godel\ to a causal outside geometry.  None of the consistent 
stringy resolutions is fully homogeneous in spacetime; in particular, space 
translations (and some of the formal supersymmetry) have been broken 
spontaneously by the domain wall.   

We would like to finish by remarking on two striking philosophical elements
which have made themselves apparent during the work described in this paper.  
The first of these is the different role that D-brane domain walls play in 
the duality between open and closed strings, in contrast to the more studied 
case of higher codimension D-branes.  
D-brane domain walls seem to exactly straddle this duality: for any coupling 
we always have both the D-brane with open strings and
a curved space with closed strings.  It would be very interesting to gain 
a more generic understanding of this phenomenon which would encompass all 
D-brane domain walls, from enhan\c{c}ons to D8-branes.  

Second, we would like to spotlight the fact that supertubes (and in some 
instances more general objects that we referred to as hypertubes) with large 
charges carry angular momentum much more effectively then their constituents.  
This in turn implies that the bound-state extended object supports a given 
amount of angular momentum in a spatial region that is much more compact 
than the region that would be effectively occupied by a collection of 
the individual unbounded constituents carrying the same angular momentum.  
Perhaps this behavior is generic in string theory, and should be 
expect of more general extended objects, even neutral black ones.

\section*{Acknowledgements}
We would like to thank
 O.~Aharony,
 R.~Bousso,
 V.~Balasubramanian, 
 J.~de~Boer,
 R.~Emparan,
 O.~Ganor,
 A.~Hashimoto,
 M.~Rangamani,
 D.~Marolf,
 D.~Minic,
 S.J.~Rey,
 and M.~Strominger
for insightful discussions. The work of EGG is
supported by the Berkeley Center for Theoretical Physics and also
partial support from the DOE grant DE-AC03-76SF00098 and the NSF
grant PHY-0098840.  The work of PH is supported by the
Berkeley Center for Theoretical Physics, NSF Grant PHY-0244900 and
DOE grant DE-AC03-76SF00098.

\newpage



\bibliography{GodelBMPV}

\providecommand{\href}[2]{#2}\begingroup\raggedright\begin{thebibliography}{10}

\bibitem{BGHV}
E.~K. Boyda, S.~Ganguli, P.~Ho\v{r}ava, and U.~Varadarajan, ``Holographic
  protection of chronology in universes of the G{\oo}del type,'' {\em Phys.
  Rev.} {\bf D67} (2003) 106003,
\href{http://www.arXiv.org/abs/hep-th/0212087}{{\tt hep-th/0212087}}.

\bibitem{GGHPR}
J.~P. Gauntlett, J.~B. Gutowski, C.~M. Hull, S.~Pakis, and H.~S. Reall, ``All
  supersymmetric solutions of minimal supergravity in five dimensions,''
\href{http://www.arXiv.org/abs/hep-th/0209114}{{\tt hep-th/0209114}}.

\bibitem{Boussoscreen}
R.~Bousso, ``The holographic principle for general backgrounds,'' {\em Class.
  Quant. Grav.} {\bf 17} (2000) 997--1005,
\href{http://www.arXiv.org/abs/hep-th/9911002}{{\tt hep-th/9911002}}.

\bibitem{3Dgodel}
N.~Drukker, B.~Fiol, and J.~Simon, ``Goedel's universe in a supertube shroud,''
  {\em Phys. Rev. Lett.} {\bf 91} (2003) 231601,
\href{http://www.arXiv.org/abs/hep-th/0306057}{{\tt hep-th/0306057}}.

\bibitem{Supertube1}
D.~Mateos and P.~K. Townsend, ``Supertubes,'' {\em Phys. Rev. Lett.} {\bf 87}
  (2001) 011602,
\href{http://www.arXiv.org/abs/hep-th/0103030}{{\tt hep-th/0103030}}.

\bibitem{Supertube2}
R.~Emparan, D.~Mateos, and P.~K. Townsend, ``Supergravity supertubes,'' {\em
  JHEP} {\bf 07} (2001) 011,
\href{http://www.arXiv.org/abs/hep-th/0106012}{{\tt hep-th/0106012}}.

\bibitem{Herdeiro}
C.~A.~R. Herdeiro, ``Spinning deformations of the D1-D5 system and a geometric
  resolution of closed timelike curves,''
\href{http://www.arXiv.org/abs/hep-th/0212002}{{\tt hep-th/0212002}}.

\bibitem{GH}
E.~G. Gimon and A.~Hashimoto, ``Black holes in G{\oo}del universes and
  pp-waves,''
\href{http://www.arXiv.org/abs/hep-th/0304181}{{\tt hep-th/0304181}}.

\bibitem{GodelBH2}
E.~G. Gimon, A.~Hashimoto, V.~E. Hubeny, O.~Lunin, and M.~Rangamani, ``Black
  strings in asymptotically plane wave geometries,'' {\em JHEP} {\bf 08} (2003)
  035,
\href{http://www.arXiv.org/abs/hep-th/0306131}{{\tt hep-th/0306131}}.

\bibitem{BDGO}
D.~Brecher, U.~H. Danielsson, J.~P. Gregory, and M.~E. Olsson, ``Rotating black
  holes in a Goedel universe,'' {\em JHEP} {\bf 11} (2003) 033,
\href{http://www.arXiv.org/abs/hep-th/0309058}{{\tt hep-th/0309058}}.

\bibitem{BD}
K.~Behrndt and D.~Klemm, ``Black holes in Goedel-type universes with a
  cosmological constant,''
\href{http://www.arXiv.org/abs/hep-th/0401239}{{\tt hep-th/0401239}}.

\bibitem{BMPV1}
J.~C. Breckenridge, R.~C. Myers, A.~W. Peet, and C.~Vafa, ``D-branes and
  spinning black holes,'' {\em Phys. Lett.} {\bf B391} (1997) 93--98,
\href{http://www.arXiv.org/abs/hep-th/9602065}{{\tt hep-th/9602065}}.

\bibitem{BMPV2}
J.~C. Breckenridge {\em et al.}, ``Macroscopic and Microscopic Entropy of
  Near-Extremal Spinning Black Holes,'' {\em Phys. Lett.} {\bf B381} (1996)
  423--426,
\href{http://www.arXiv.org/abs/hep-th/9603078}{{\tt hep-th/9603078}}.

\bibitem{CveticYoum}
M.~Cvetic and D.~Youm, ``General Rotating Five Dimensional Black Holes of
  Toroidally Compactified Heterotic String,'' {\em Nucl. Phys.} {\bf B476}
  (1996) 118--132,
\href{http://www.arXiv.org/abs/hep-th/9603100}{{\tt hep-th/9603100}}.

\bibitem{CveticLarsen}
M.~Cvetic and F.~Larsen, ``Near horizon geometry of rotating black holes in
  five dimensions,'' {\em Nucl. Phys.} {\bf B531} (1998) 239--255,
\href{http://www.arXiv.org/abs/hep-th/9805097}{{\tt hep-th/9805097}}.

\bibitem{Drukker}
N.~Drukker, ``Supertube domain-walls and elimination of closed time-like curves
  in string theory,''
\href{http://www.arXiv.org/abs/hep-th/0404239}{{\tt hep-th/0404239}}.

\bibitem{GMT}
J.~P. Gauntlett, R.~C. Myers, and P.~K. Townsend, ``Supersymmetry of rotating
  branes,'' {\em Phys. Rev.} {\bf D59} (1999) 025001,
\href{http://www.arXiv.org/abs/hep-th/9809065}{{\tt hep-th/9809065}}.

\bibitem{HT}
T.~Harmark and T.~Takayanagi, ``Supersymmetric Goedel universes in string
  theory,'' {\em Nucl. Phys.} {\bf B662} (2003) 3--39,
\href{http://www.arXiv.org/abs/hep-th/0301206}{{\tt hep-th/0301206}}.

\bibitem{clp}
M.~Cvetic, H.~Lu, and C.~N. Pope, ``Penrose limits, pp-waves and deformed
  M2-branes,'' {\em Phys. Rev.} {\bf D69} (2004) 046003,
\href{http://www.arXiv.org/abs/hep-th/0203082}{{\tt hep-th/0203082}}.

\bibitem{ferrara}
S.~Ferrara, R.~Kallosh, and A.~Strominger, ``N=2 extremal black holes,'' {\em
  Phys. Rev.} {\bf D52} (1995) 5412--5416,
\href{http://www.arXiv.org/abs/hep-th/9508072}{{\tt hep-th/9508072}}.

\bibitem{Gibbons}
G.~W. Gibbons and C.~A.~R. Herdeiro, ``Supersymmetric rotating black holes and
  causality violation,'' {\em Class. Quant. Grav.} {\bf 16} (1999) 3619--3652,
\href{http://www.arXiv.org/abs/hep-th/9906098}{{\tt hep-th/9906098}}.

\bibitem{Maldacena}
J.~M. Maldacena, ``The large N limit of superconformal field theories and
  supergravity,'' {\em Adv. Theor. Math. Phys.} {\bf 2} (1998) 231--252,
\href{http://www.arXiv.org/abs/hep-th/9711200}{{\tt hep-th/9711200}}.

\bibitem{BK}
I.~Bena and P.~Kraus, ``Three charge supertubes and black hole hair,''
\href{http://www.arXiv.org/abs/hep-th/0402144}{{\tt hep-th/0402144}}.

\bibitem{Mathur}
S.~D. Mathur, ``Gravity on AdS(3) and flat connections in the boundary CFT,''
\href{http://www.arXiv.org/abs/hep-th/0101118}{{\tt hep-th/0101118}}.

\bibitem{MS}
J.~M. Maldacena and L.~Susskind, ``D-branes and Fat Black Holes,'' {\em Nucl.
  Phys.} {\bf B475} (1996) 679--690,
\href{http://www.arXiv.org/abs/hep-th/9604042}{{\tt hep-th/9604042}}.

\bibitem{Supertube3}
D.~Mateos, S.~Ng, and P.~K. Townsend, ``Tachyons, supertubes and
  brane/anti-brane systems,'' {\em JHEP} {\bf 03} (2002) 016,
\href{http://www.arXiv.org/abs/hep-th/0112054}{{\tt hep-th/0112054}}.

\bibitem{Supercurves}
D.~Mateos, S.~Ng, and P.~K. Townsend, ``Supercurves,'' {\em Phys. Lett.} {\bf
  B538} (2002) 366--374,
\href{http://www.arXiv.org/abs/hep-th/0204062}{{\tt hep-th/0204062}}.

\bibitem{PM}
B.~C. Palmer and D.~Marolf, ``Counting supertubes,''
\href{http://www.arXiv.org/abs/hep-th/0403025}{{\tt hep-th/0403025}}.

\bibitem{ML1}
O.~Lunin and S.~D. Mathur, ``Metric of the multiply wound rotating string,''
  {\em Nucl. Phys.} {\bf B610} (2001) 49--76,
\href{http://www.arXiv.org/abs/hep-th/0105136}{{\tt hep-th/0105136}}.

\bibitem{ML2}
O.~Lunin and S.~D. Mathur, ``The slowly rotating near extremal D1-D5 system as
  a 'hot tube','' {\em Nucl. Phys.} {\bf B615} (2001) 285--312,
\href{http://www.arXiv.org/abs/hep-th/0107113}{{\tt hep-th/0107113}}.

\bibitem{ML3}
O.~Lunin and S.~D. Mathur, ``AdS/CFT duality and the black hole information
  paradox,'' {\em Nucl. Phys.} {\bf B623} (2002) 342--394,
\href{http://www.arXiv.org/abs/hep-th/0109154}{{\tt hep-th/0109154}}.

\bibitem{ML4}
O.~Lunin and S.~D. Mathur, ``Statistical interpretation of Bekenstein entropy
  for systems with a stretched horizon,'' {\em Phys. Rev. Lett.} {\bf 88}
  (2002) 211303,
\href{http://www.arXiv.org/abs/hep-th/0202072}{{\tt hep-th/0202072}}.

\bibitem{ML5}
O.~Lunin, S.~D. Mathur, and A.~Saxena, ``What is the gravity dual of a chiral
  primary?,'' {\em Nucl. Phys.} {\bf B655} (2003) 185--217,
\href{http://www.arXiv.org/abs/hep-th/0211292}{{\tt hep-th/0211292}}.

\bibitem{ML6}
O.~Lunin, J.~Maldacena, and L.~Maoz, ``Gravity solutions for the D1-D5 system
  with angular momentum,''
\href{http://www.arXiv.org/abs/hep-th/0212210}{{\tt hep-th/0212210}}.

\bibitem{Dyson}
L.~Dyson, ``Chronology protection in string theory,'' {\em JHEP} {\bf 03}
  (2004) 024,
\href{http://www.arXiv.org/abs/hep-th/0302052}{{\tt hep-th/0302052}}.

\bibitem{Israel}
D.~Israel, ``Quantization of heterotic strings in a Goedel/anti de Sitter
  spacetime and chronology protection,'' {\em JHEP} {\bf 01} (2004) 042,
\href{http://www.arXiv.org/abs/hep-th/0310158}{{\tt hep-th/0310158}}.

\bibitem{Lunin}
O.~Lunin, ``Adding momentum to D1-D5 system,''
\href{http://www.arXiv.org/abs/hep-th/0404006}{{\tt hep-th/0404006}}.

\bibitem{BL}
D.~Bak and K.-M. Lee, ``Noncommutative supersymmetric tubes,'' {\em Phys.
  Lett.} {\bf B509} (2001) 168--174,
\href{http://www.arXiv.org/abs/hep-th/0103148}{{\tt hep-th/0103148}}.

\bibitem{SWnc}
N.~Seiberg and E.~Witten, ``String theory and noncommutative geometry,'' {\em
  JHEP} {\bf 09} (1999) 032,
\href{http://www.arXiv.org/abs/hep-th/9908142}{{\tt hep-th/9908142}}.

\bibitem{Aspects}
M.~Kruczenski, R.~C. Myers, A.~W. Peet, and D.~J. Winters, ``Aspects of
  supertubes,'' {\em JHEP} {\bf 05} (2002) 017,
\href{http://www.arXiv.org/abs/hep-th/0204103}{{\tt hep-th/0204103}}.

\bibitem{Takayanagi}
H.~Takayanagi, ``Boundary states for supertubes in flat spacetime and Goedel
  universe,'' {\em JHEP} {\bf 12} (2003) 011,
\href{http://www.arXiv.org/abs/hep-th/0309135}{{\tt hep-th/0309135}}.

\end{thebibliography}\endgroup
\bibliographystyle{utphys}
\end{document}